%% file: archive-main.tex
\PassOptionsToPackage{dvipsnames,table}{xcolor}
\documentclass[pdflatex,sn-mathphys-num]{sn-jnl}%

\usepackage{tcolorbox}
\usepackage{tikz}
\input{templates/general}

\input{templates/kit-colors}
\input{templates/listings}
\input{templates/tikz-vitruvius}

\usepackage[nohyperlinks,printonlyused,nolist]{acronym}
\usepackage{xspace}

\usepackage{graphicx}%
\usepackage{multirow}%
\usepackage{amsmath,amssymb,amsfonts}%
\usepackage{amsthm}%
\usepackage{mathrsfs}%
\usepackage[title]{appendix}%
\usepackage{xcolor}%
\usepackage{textcomp}%
\usepackage{manyfoot}%
\usepackage{booktabs}%
\usepackage{algorithm}%
\usepackage{algorithmicx}%
\usepackage{algpseudocode}%
\usepackage{listings}%

\usepackage{tikz}
\usetikzlibrary{shapes, arrows, positioning, fit, calc}
\usepackage{hyperref}
\usepackage{xspace}
\usepackage{cleveref}
\usepackage{listings}
\usepackage{xcolor}
\usepackage{graphicx}
\usepackage{booktabs}
\usepackage{tabularx}
\usepackage{multirow}
\usepackage{bm}

\usepackage{amsmath}
\usepackage{amssymb}
\usepackage{amsthm}
\usepackage{listings}
\usepackage{xcolor}
\usepackage{cleveref}
\usepackage{graphicx}

\usepackage{enumitem}
\setlist{nosep, leftmargin=*}
\usepackage{tikz}
\usetikzlibrary{positioning,arrows.meta,calc}
\usetikzlibrary{arrows.meta}
\usepackage{bm}
\usepackage{dblfloatfix}
\usetikzlibrary{positioning,arrows.meta,fit}
\usetikzlibrary{fit,positioning,arrows.meta,shapes.multipart}
\usetikzlibrary{shapes.multipart}
\usepackage[textsize=tiny,backgroundcolor=yellow!30]{todonotes}

\newtheorem{assumption}{Assumption}
\newtheorem{definition}{Definition}
\newtheorem{proposition}{Proposition}

\providecommand{\Description}[1]{%
\par\small \textbf{Description: }#1 }

\newcommand{\vsum}{%
\textsc{%
V\kern-1pt%
-\kern-.5pt%
S\kern-.5pt%
U\kern-.5pt%
M\kern-.5pt%
}\xspace}

\theoremstyle{thmstyleone}%

\theoremstyle{thmstyletwo}%

\theoremstyle{thmstylethree}%

\begin{document}

\title[Replay-Based Merge for Delta-Based Consistency Management in  Multi-Model Development]{Replay-Based Merge for Delta-Based Consistency Management in  Multi-Model Development}

\author*[1]{\fnm{Razieh} \sur{Dehghani}}\email{raziyeh.dehghani@kit.edu}

\author[1]{\fnm{Arne} \sur{Lange}}\email{arne.lange@kit.edu}

\author[1]{\fnm{Martin} \sur{Armbruster}}\email{martin.armbruster@kit.edu}

\author[1]{\fnm{Linh} \sur{Nguyen Phuong}}\email{linh.phuong@student.kit.edu}

\author[2]{\fnm{Thomas} \sur{Alexander Voelk}}\email{thomas.voelk@kit.edu}

\author[1]{\fnm{Benedikt} \sur{Jutz}}\email{benedikt.jutz@kit.edu}

\author[1]{\fnm{Miriam} \sur{Boss}}\email{boss.miriam@kit.edu}

\author[2]{\fnm{Albert} \sur{Albers}}\email{albert.albers@kit.edu}

\author[2]{\fnm{Tobias} \sur{Düser}}\email{tobias.dueser@kit.edu}

\author[1]{\fnm{Ralf} \sur{Reussner}}\email{reussner@kit.edu}

\author[1]{\fnm{Anne} \sur{Koziolek}}\email{koziolek@kit.edu}

\affil*[1]{\orgdiv{Institute of Information Security and Dependability (KASTEL)}, \orgname{Karlsruhe Institute of Technology (KIT)}, \orgaddress{\street{Am Fasanengarten 5}, \city{Karlsruhe}, \postcode{76131}, \state{Baden-Württemberg}, \country{Germany}}}

\affil[2]{\orgdiv{Institute of Product Development (IPEK)}, \orgname{Karlsruhe Institute of Technology (KIT)}, \orgaddress{\street{Kaiserstr. 10}, \city{Karlsruhe}, \postcode{76131}, \state{Baden-Württemberg}, \country{Germany}}}

\abstract{

When multiple models are kept consistent by automated change propagation rules, merging independently evolved branches poses a challenge that existing model merge approaches cannot fully address.
While operation-based versioning tools record edit histories, they do not account for multi-model consistency rules that generate consequential changes across model boundaries. This leads to false-positive conflicts on consequential model elements that would be regenerated during a consistency-aware merge.

We present a replay-based three-way merge for multi-model consistency management approaches that distinguish original from consequential changes. Our approach records original changes per commit and interleaves them from both branches through the consistency preservation engine, using a dependency graph to find an ordering that preserves both branches' original intent while consequential changes are regenerated rather than merged. Conflicts are classified by provenance: direct conflicts between original changes require human resolution, and consequential overlaps become ordering constraints that the algorithm resolves automatically; a cycle in the dependency graph indicates a consequential conflict. We instantiate and evaluate the approach in the \vitruv framework for view-based software development.

We formalize the merge operator over change histories and show that every successful merge preserves global consistency. We implement and evaluate the operator in the \vitruv multi-model management framework. We evaluate on three three-model case studies, two from automotive domains, maintaining value-propagation consistency relations, and one from software engineering, maintaining structural ones, and compare against EMF~Compare as baseline: across 14 scenarios, our approach reports 10 blocking conflicts while EMF~Compare reports 37. The difference arises because EMF~Compare cannot distinguish original from consequential changes and therefore reports conflicts on consequential model elements that our approach handles automatically.

}

\keywords{Model Consistency, Three-Way Merge, Change Propagation, Model Versioning, Conflict Detection}

\maketitle

\input{templates/acronyms}

\input{sections/01_introduction}
\input{sections/02_foundation}
\input{sections/03_illustratingexample}
\input{sections/04_formalization}
\input{sections/05_implementation}
\input{sections/06_evaluation}
\input{sections/07_relatedwork}
\input{sections/08_conclusion}

\bmhead{Acknowledgements}

This work was supported by the DFG (German Research Foundation) with the Collaborative Research Center “Convide” — SFB 1608 — 501798263 and funded by the topic Engineering Secure Systems, KASTEL Security Research Labs by the Helmholtz Association (HGF).

\bibliography{archive-refs}%

\end{document}

%% file: templates/general.tex
\usepackage[T1]{fontenc}
\usepackage[scale=.9]{FiraMono}

\usepackage{amsmath,amssymb,amsfonts}
\usepackage{graphicx}
\usepackage{textcomp}
\usepackage[dvipsnames,table]{xcolor}
\usepackage[inline]{enumitem}

\usepackage{xspace}
\hypersetup{
    pdfborder={0 0 0}
}
\usepackage[nameinlink]{cleveref}
\makeatletter
\AtBeginDocument
 {
   \def\ltx@label#1{\cref@label{#1}}
   \def\label@in@display@noarg#1{\cref@old@label@in@display{#1}}%
\def\label@in@mmeasure@noarg#1{
    \begingroup%
      \measuring@false%
      \cref@old@label@in@display{#1}
    \endgroup}%
 }
\makeatother

\usepackage{csquotes}
\usepackage{svg}
\usepackage{float}
\newcommand{\vitruv}{%
\textsc{%
V\kern-1pt%
i\kern-.5pt%
t\kern-.5pt%
r\kern-.5pt%
u\kern-.5pt%
v\kern-.5pt%
i\kern-.5pt%
u\kern-.5pt%
s\kern-.5pt%
}\xspace}

\usetikzlibrary{calc}

\def\BibTeX{{\rm B\kern-.05em{\sc i\kern-.025em b}\kern-.08em
    T\kern-.1667em\lower.7ex\hbox{E}\kern-.125emX}}

\makeatletter
\newcommand{\linebreakand}{%
  \end{@IEEEauthorhalign}
  \hfill\mbox{}\par
  \mbox{}\hfill\begin{@IEEEauthorhalign}
}
\makeatother

\newcommand{\Metamodel}{Metamodel\xspace}

%% file: templates/kit-colors.tex
\definecolor{kit-green}{RGB}{0, 150, 130}
\definecolor{kit-green100}{RGB}{0, 150, 130}
\definecolor{kit-green90}{rgb}{0.1, 0.6294, 0.5588}
\definecolor{kit-green80}{rgb}{0.2, 0.6706, 0.6078}
\definecolor{kit-green75}{rgb}{0.25, 0.6912, 0.6324}
\definecolor{kit-green70}{rgb}{0.3, 0.7118, 0.6569}
\definecolor{kit-green60}{rgb}{0.4, 0.7529, 0.7059}
\definecolor{kit-green50}{rgb}{0.5, 0.7941, 0.7549}
\definecolor{kit-green40}{rgb}{0.6, 0.8353, 0.8039}
\definecolor{kit-green30}{rgb}{0.7, 0.8765, 0.8529}
\definecolor{kit-green25}{rgb}{0.75, 0.8971, 0.8775}
\definecolor{kit-green20}{rgb}{0.8, 0.9176, 0.902}
\definecolor{kit-green15}{rgb}{0.85, 0.9382, 0.9265}
\definecolor{kit-green10}{rgb}{0.9, 0.9588, 0.951}
\definecolor{kit-green5}{rgb}{0.95, 0.9794, 0.9755}

\definecolor{kit-blue}{RGB}{70, 100, 170}
\definecolor{kit-blue100}{RGB}{70, 100, 170}
\definecolor{kit-blue90}{rgb}{0.3471, 0.4529, 0.7}
\definecolor{kit-blue80}{rgb}{0.4196, 0.5137, 0.7333}
\definecolor{kit-blue75}{rgb}{0.4559, 0.5441, 0.75}
\definecolor{kit-blue70}{rgb}{0.4922, 0.5745, 0.7667}
\definecolor{kit-blue60}{rgb}{0.5647, 0.6353, 0.8}
\definecolor{kit-blue50}{rgb}{0.6373, 0.6961, 0.8333}
\definecolor{kit-blue40}{rgb}{0.7098, 0.7569, 0.8667}
\definecolor{kit-blue30}{rgb}{0.7824, 0.8176, 0.9}
\definecolor{kit-blue25}{rgb}{0.8186, 0.848, 0.9167}
\definecolor{kit-blue20}{rgb}{0.8549, 0.8784, 0.9333}
\definecolor{kit-blue15}{rgb}{0.8912, 0.9088, 0.95}
\definecolor{kit-blue10}{rgb}{0.9275, 0.9392, 0.9667}
\definecolor{kit-blue5}{rgb}{0.9637, 0.9696, 0.9833}

\definecolor{kit-red}{RGB}{162, 34, 35}
\definecolor{kit-red100}{RGB}{162, 34, 35}
\definecolor{kit-red90}{rgb}{0.6718, 0.22, 0.2235}
\definecolor{kit-red80}{rgb}{0.7082, 0.3067, 0.3098}
\definecolor{kit-red75}{rgb}{0.7265, 0.35, 0.3529}
\definecolor{kit-red70}{rgb}{0.7447, 0.3933, 0.3961}
\definecolor{kit-red60}{rgb}{0.7812, 0.48, 0.4824}
\definecolor{kit-red50}{rgb}{0.8176, 0.5667, 0.5686}
\definecolor{kit-red40}{rgb}{0.8541, 0.6533, 0.6549}
\definecolor{kit-red30}{rgb}{0.8906, 0.74, 0.7412}
\definecolor{kit-red25}{rgb}{0.9088, 0.7833, 0.7843}
\definecolor{kit-red20}{rgb}{0.9271, 0.8267, 0.8275}
\definecolor{kit-red15}{rgb}{0.9453, 0.87, 0.8706}
\definecolor{kit-red10}{rgb}{0.9635, 0.9133, 0.9137}
\definecolor{kit-red5}{rgb}{0.9818, 0.9567, 0.9569}

\definecolor{kit-yellow}{RGB}{252, 229, 0}
\definecolor{kit-yellow100}{RGB}{252, 229, 0}
\definecolor{kit-yellow90}{rgb}{0.9894, 0.9082, 0.1}
\definecolor{kit-yellow80}{rgb}{0.9906, 0.9184, 0.2}
\definecolor{kit-yellow75}{rgb}{0.9912, 0.9235, 0.25}
\definecolor{kit-yellow70}{rgb}{0.9918, 0.9286, 0.3}
\definecolor{kit-yellow60}{rgb}{0.9929, 0.9388, 0.4}
\definecolor{kit-yellow50}{rgb}{0.9941, 0.949, 0.5}
\definecolor{kit-yellow40}{rgb}{0.9953, 0.9592, 0.6}
\definecolor{kit-yellow30}{rgb}{0.9965, 0.9694, 0.7}
\definecolor{kit-yellow25}{rgb}{0.9971, 0.9745, 0.75}
\definecolor{kit-yellow20}{rgb}{0.9976, 0.9796, 0.8}
\definecolor{kit-yellow15}{rgb}{0.9982, 0.9847, 0.85}
\definecolor{kit-yellow10}{rgb}{0.9988, 0.9898, 0.9}
\definecolor{kit-yellow5}{rgb}{0.9994, 0.9949, 0.95}

\definecolor{kit-orange}{RGB}{223, 155, 27}
\definecolor{kit-orange100}{RGB}{223, 155, 27}
\definecolor{kit-orange90}{rgb}{0.8871, 0.6471, 0.1953}
\definecolor{kit-orange80}{rgb}{0.8996, 0.6863, 0.2847}
\definecolor{kit-orange75}{rgb}{0.9059, 0.7059, 0.3294}
\definecolor{kit-orange70}{rgb}{0.9122, 0.7255, 0.3741}
\definecolor{kit-orange60}{rgb}{0.9247, 0.7647, 0.4635}
\definecolor{kit-orange50}{rgb}{0.9373, 0.8039, 0.5529}
\definecolor{kit-orange40}{rgb}{0.9498, 0.8431, 0.6424}
\definecolor{kit-orange30}{rgb}{0.9624, 0.8824, 0.7318}
\definecolor{kit-orange25}{rgb}{0.9686, 0.902, 0.7765}
\definecolor{kit-orange20}{rgb}{0.9749, 0.9216, 0.8212}
\definecolor{kit-orange15}{rgb}{0.9812, 0.9412, 0.8659}
\definecolor{kit-orange10}{rgb}{0.9875, 0.9608, 0.9106}
\definecolor{kit-orange5}{rgb}{0.9937, 0.9804, 0.9553}

\definecolor{kit-lightgreen}{RGB}{140, 182, 60}
\definecolor{kit-lightgreen100}{RGB}{140, 182, 60}
\definecolor{kit-lightgreen90}{rgb}{0.5941, 0.7424, 0.3118}
\definecolor{kit-lightgreen80}{rgb}{0.6392, 0.771, 0.3882}
\definecolor{kit-lightgreen75}{rgb}{0.6618, 0.7853, 0.4265}
\definecolor{kit-lightgreen70}{rgb}{0.6843, 0.7996, 0.4647}
\definecolor{kit-lightgreen60}{rgb}{0.7294, 0.8282, 0.5412}
\definecolor{kit-lightgreen50}{rgb}{0.7745, 0.8569, 0.6176}
\definecolor{kit-lightgreen40}{rgb}{0.8196, 0.8855, 0.6941}
\definecolor{kit-lightgreen30}{rgb}{0.8647, 0.9141, 0.7706}
\definecolor{kit-lightgreen25}{rgb}{0.8873, 0.9284, 0.8088}
\definecolor{kit-lightgreen20}{rgb}{0.9098, 0.9427, 0.8471}
\definecolor{kit-lightgreen15}{rgb}{0.9324, 0.9571, 0.8853}
\definecolor{kit-lightgreen10}{rgb}{0.9549, 0.9714, 0.9235}
\definecolor{kit-lightgreen5}{rgb}{0.9775, 0.9857, 0.9618}

\definecolor{kit-purple}{RGB}{163, 16, 124}
\definecolor{kit-purple100}{RGB}{163, 16, 124}
\definecolor{kit-purple90}{rgb}{0.6753, 0.1565, 0.5376}
\definecolor{kit-purple80}{rgb}{0.7114, 0.2502, 0.589}
\definecolor{kit-purple75}{rgb}{0.7294, 0.2971, 0.6147}
\definecolor{kit-purple70}{rgb}{0.7475, 0.3439, 0.6404}
\definecolor{kit-purple60}{rgb}{0.7835, 0.4376, 0.6918}
\definecolor{kit-purple50}{rgb}{0.8196, 0.5314, 0.7431}
\definecolor{kit-purple40}{rgb}{0.8557, 0.6251, 0.7945}
\definecolor{kit-purple30}{rgb}{0.8918, 0.7188, 0.8459}
\definecolor{kit-purple25}{rgb}{0.9098, 0.7657, 0.8716}
\definecolor{kit-purple20}{rgb}{0.9278, 0.8125, 0.8973}
\definecolor{kit-purple15}{rgb}{0.9459, 0.8594, 0.9229}
\definecolor{kit-purple10}{rgb}{0.9639, 0.9063, 0.9486}
\definecolor{kit-purple5}{rgb}{0.982, 0.9531, 0.9743}

\definecolor{kit-brown}{RGB}{167, 130, 46}
\definecolor{kit-brown100}{RGB}{167, 130, 46}
\definecolor{kit-brown90}{rgb}{0.6894, 0.5588, 0.2624}
\definecolor{kit-brown80}{rgb}{0.7239, 0.6078, 0.3443}
\definecolor{kit-brown75}{rgb}{0.7412, 0.6324, 0.3853}
\definecolor{kit-brown70}{rgb}{0.7584, 0.6569, 0.4263}
\definecolor{kit-brown60}{rgb}{0.7929, 0.7059, 0.5082}
\definecolor{kit-brown50}{rgb}{0.8275, 0.7549, 0.5902}
\definecolor{kit-brown40}{rgb}{0.862, 0.8039, 0.6722}
\definecolor{kit-brown30}{rgb}{0.8965, 0.8529, 0.7541}
\definecolor{kit-brown25}{rgb}{0.9137, 0.8775, 0.7951}
\definecolor{kit-brown20}{rgb}{0.931, 0.902, 0.8361}
\definecolor{kit-brown15}{rgb}{0.9482, 0.9265, 0.8771}
\definecolor{kit-brown10}{rgb}{0.9655, 0.951, 0.918}
\definecolor{kit-brown5}{rgb}{0.9827, 0.9755, 0.959}

\definecolor{kit-cyan}{RGB}{35, 161, 224}
\definecolor{kit-cyan100}{RGB}{35, 161, 224}
\definecolor{kit-cyan90}{rgb}{0.2235, 0.6682, 0.8906}
\definecolor{kit-cyan80}{rgb}{0.3098, 0.7051, 0.9027}
\definecolor{kit-cyan75}{rgb}{0.3529, 0.7235, 0.9088}
\definecolor{kit-cyan70}{rgb}{0.3961, 0.742, 0.9149}
\definecolor{kit-cyan60}{rgb}{0.4824, 0.7788, 0.9271}
\definecolor{kit-cyan50}{rgb}{0.5686, 0.8157, 0.9392}
\definecolor{kit-cyan40}{rgb}{0.6549, 0.8525, 0.9514}
\definecolor{kit-cyan30}{rgb}{0.7412, 0.8894, 0.9635}
\definecolor{kit-cyan25}{rgb}{0.7843, 0.9078, 0.9696}
\definecolor{kit-cyan20}{rgb}{0.8275, 0.9263, 0.9757}
\definecolor{kit-cyan15}{rgb}{0.8706, 0.9447, 0.9818}
\definecolor{kit-cyan10}{rgb}{0.9137, 0.9631, 0.9878}
\definecolor{kit-cyan5}{rgb}{0.9569, 0.9816, 0.9939}

\definecolor{kit-gray}{RGB}{0, 0, 0}
\definecolor{kit-gray100}{RGB}{0, 0, 0}
\definecolor{kit-gray90}{rgb}{0.1, 0.1, 0.1}
\definecolor{kit-gray80}{rgb}{0.2, 0.2, 0.2}
\definecolor{kit-gray75}{rgb}{0.25, 0.25, 0.25}
\definecolor{kit-gray70}{rgb}{0.3, 0.3, 0.3}
\definecolor{kit-gray60}{rgb}{0.4, 0.4, 0.4}
\definecolor{kit-gray50}{rgb}{0.5, 0.5, 0.5}
\definecolor{kit-gray40}{rgb}{0.6, 0.6, 0.6}
\definecolor{kit-gray30}{rgb}{0.7, 0.7, 0.7}
\definecolor{kit-gray25}{rgb}{0.75, 0.75, 0.75}
\definecolor{kit-gray20}{rgb}{0.8, 0.8, 0.8}
\definecolor{kit-gray15}{rgb}{0.85, 0.85, 0.85}
\definecolor{kit-gray10}{rgb}{0.9, 0.9, 0.9}
\definecolor{kit-gray5}{rgb}{0.95, 0.95, 0.95}

\definecolor{researchareaAcolor}{named}{kit-orange50}
\definecolor{researchareaBcolor}{named}{kit-blue50}
\definecolor{researchareaCcolor}{named}{kit-lightgreen50}
\definecolor{researchareaZcolor}{named}{white}

%% file: templates/listings.tex
\usepackage{listings}
\definecolor{backcolor}{RGB}{226, 223, 223}

\lstdefinelanguage{reactions}{
    alsodigit = {},
    keywords=[1]{
    import, as, reactions, in, reaction, to, changes, in, execute, actions, after, attribute, replaced, at, call, routine, match, update, retrieve, corresponding, val, create, element, created, and, inserted, new, with
    },
    xleftmargin=2.5em,
    keywordstyle=[1]\color{Mahogany},
    sensitive=true,
    stringstyle=\color{red},
    numbers=left,
    stepnumber=1,
    showstringspaces=false,
    tabsize=1,
    breaklines=true,
    breakatwhitespace=false,
    basicstyle=\scriptsize\ttfamily,
    print,
    backgroundcolor=\color{backcolor},
    escapechar=@
}

\lstdefinelanguage{Reactions2}[]{reactions}{
    numbers=none
}

\lstdefinelanguage{neojoin}{
    alsodigit = {},
    keywords=[1]{
        export, to, import, as, from, where, create
    },
    morestring=[b]",
    xleftmargin=2.5em,
    keywordstyle=[1]\color{Mahogany},
    sensitive=true,
    stringstyle=\color{red},
    numbers=left,
    stepnumber=1,
    showstringspaces=false,
    tabsize=1,
    breaklines=true,
    breakatwhitespace=false,
    basicstyle=\scriptsize\ttfamily,
    print,
    backgroundcolor=\color{backcolor},
    escapechar=@
}

\lstdefinelanguage{javax}[]{Java}{
    alsodigit = {},
    morekeywords = {var},
    xleftmargin=2.5em,
    keywordstyle=[1]\color{Mahogany},
    commentstyle=\color{PineGreen},
    sensitive=true,
    stringstyle=\color{CornflowerBlue},
    numbers=left,
    stepnumber=1,
    showstringspaces=false,
    tabsize=1,
    breaklines=true,
    breakatwhitespace=false,
    basicstyle=\scriptsize\ttfamily,
    print,
    backgroundcolor=\color{backcolor},
}
\crefname{lstlisting}{listing}{listings}
\Crefname{lstlisting}{Listing}{Listings}

%% file: templates/tikz-vitruvius.tex
\usetikzlibrary{positioning}
\usetikzlibrary{arrows}
\usetikzlibrary{arrows.meta}
\usetikzlibrary{bending}
\usetikzlibrary{fit}
\usetikzlibrary{petri}
\usetikzlibrary{decorations.markings}
\usetikzlibrary{decorations.pathmorphing}
\usetikzlibrary{decorations.pathreplacing}
\usetikzlibrary{decorations.text}
\usetikzlibrary{fit}
\usetikzlibrary{automata}
\usetikzlibrary{shapes}
\usetikzlibrary{shadows}
\usetikzlibrary{calc}
\usetikzlibrary{patterns}

\pgfdeclareshape{circle vertical split}
{
  \inheritsavedanchors[from=circle]
  \inheritanchorborder[from=circle]
  \inheritanchor[from=circle]{north}
  \inheritanchor[from=circle]{north west}
  \inheritanchor[from=circle]{north east}
  \inheritanchor[from=circle]{center}
  \inheritanchor[from=circle]{west}
  \inheritanchor[from=circle]{east}
  \inheritanchor[from=circle]{mid}
  \inheritanchor[from=circle]{mid west}
  \inheritanchor[from=circle]{mid east}
  \inheritanchor[from=circle]{base}
  \inheritanchor[from=circle]{base west}
  \inheritanchor[from=circle]{base east}
  \inheritanchor[from=circle]{south}
  \inheritanchor[from=circle]{south west}
  \inheritanchor[from=circle]{south east}
  \inheritbackgroundpath[from=circle]
  \foregroundpath{
    \centerpoint%
    \pgf@xc=\pgf@x%
    \pgf@yc=\pgf@y%
    \pgfutil@tempdima=\radius%
    \pgfmathsetlength{\pgf@xb}{\pgfkeysvalueof{/pgf/outer xsep}}%
    \pgfmathsetlength{\pgf@yb}{\pgfkeysvalueof{/pgf/outer ysep}}%
    \ifdim\pgf@xb<\pgf@yb%
      \advance\pgfutil@tempdima by-\pgf@yb%
    \else%
      \advance\pgfutil@tempdima by-\pgf@xb%
    \fi%
    \advance\pgfutil@tempdima by-.5\pgflinewidth%
    \pgfsetshortenstart{0pt}%
    \pgfsetshortenend{0pt}%
    \pgfsetarrows{-}%
    \pgfpathmoveto{\pgfpointadd{\centerpoint}{\pgfqpoint{0pt}{-1\pgfutil@tempdima}}}%
    \pgfpathlineto{\pgfpointadd{\centerpoint}{\pgfqpoint{0pt}{\pgfutil@tempdima}}}%
    \pgfusepath{stroke}%
  }
}

\tikzset{
blueshading/.style={shade, top color=white, bottom color=kit-blue30, draw=kit-blue100, thick},
lightblueshading/.style={shade, top color=white, bottom color=kit-blue15, draw=kit-blue100, thick},
greyshading/.style={shade, top color=white, bottom color=black!30, draw=black!50, thick},
lightgreyshading/.style={shade, top color=white, bottom color=black!15, draw=black!50, thick},
blackshading/.style={shade, top color=black!50, bottom color=black!95, middle color=black!95, draw=black!95, thick},
greenshading/.style={shade, top color=white, bottom color=kit-green30, draw=kit-green100, thick},
lightgreenshading/.style={shade, top color=white, bottom color=kit-green15, draw=kit-green100, thick},
redshading/.style={shade, top color=white, bottom color=kit-red30, draw=kit-red100, thick},
yellowshading/.style={shade, top color=white, bottom color=kit-yellow30, draw=kit-orange100, thick},
maygreenshading/.style={shade, top color=white, bottom color=kit-lightgreen30, draw=kit-lightgreen100, thick},
orangeshading/.style={shade, top color=white, bottom color=kit-orange30, draw=kit-orange100, thick},
purpleshading/.style={shade, top color=white, bottom color=kit-purple30, draw=kit-purple100, thick},
cyanshading/.style={shade, top color=white, bottom color=kit-cyan30, draw=kit-cyan100, thick},
roundbox/.style={rectangle, rounded corners=.5ex},
seffaction/.style={rectangle split, rectangle split parts=2, rectangle split part align={center}, text centered, rounded corners},
pcmball/.style={shade, ball color=kit-blue100,circle,fill},
startaction/.style={pcmball,minimum size=1ex},
stopaction/.style={startaction,draw=black,double},
pnplace/.style={circle, draw, minimum size=2em},
pntrans/.style={rectangle, draw, minimum height=2em},
qpnplace/.style={circle vertical split, draw, minimum size=2em},
subnetplace/.style={qpnplace, double},
}

\tikzset{uml box/.style={draw, minimum width=1cm, minimum height=.7cm, text badly centered},
	uml small box/.style={draw, minimum width=.7cm, minimum height=.5cm, text badly centered},
	uml class/.style={rectangle split, rectangle split parts=2, draw, font=\sffamily, rectangle split part align={centered,left}},
	uml full class/.style={rectangle split, rectangle split parts=3, draw, font=\sffamily, rectangle split part align={center,left,left}},
	uml activity/.style={rectangle, draw, rounded corners=.5em, minimum width=.7cm, font=\sffamily, text badly centered},
	uml activity parameter/.style={rectangle, draw, minimum width=.7cm, font=\sffamily, text badly centered},
	uml activity label/.style={font=\footnotesize},
	uml use case/.style={ellipse, draw, font=\sffamily},
	uml directed/.style={thick, -angle 90},
	uml aggregation head/.style={shape=diamond, aspect=.5, anchor=north, thick, draw},
	uml composition head/.style={shape=diamond, aspect=.5, anchor=north, thick, filldraw},
	uml generalization head/.style={shape=isosceles triangle, isosceles triangle apex angle=60, shape border rotate=90, scale=.8, draw, anchor=apex},
	uml role/.style={font=\footnotesize\sffamily},
	uml role start/.style={uml role, near start, auto},
	uml role end/.style={uml role, near end, auto},
	uml cardinality start/.style={uml role, near start, auto},
	uml cardinality end/.style={uml role, near end, auto},
	uml association name/.style={uml role, midway, auto},
	instanceof/.style={-latex,densely dashed},
	uml start state/.style={circle,fill=black},
	uml final state/.style={circle,fill=black,draw=white,very thick, double=black},
	uml pin/.style={draw, fill=white, text width=1ex, minimum height=1ex, inner sep=0},
	uml fork/.style={draw, thick, fill=black, inner sep=0ex, text width=2em},
	uml diamond/.style={diamond, draw, text width=.5em},
	uml abstract/.style={font=\sffamily\itshape},
	uml lifeline/.style={draw, fill=white, text width=0em, anchor=north},
}%

%% file: templates/acronyms.tex
\begin{acronym}
    \acro{MBSE}{Model-Based Systems Engineering}
    \acro{MoM}{Model Management}
    \acro{VSUM}[V-SUM]{Virtual Single Underlying Model}
    \acro{VSUMM}[V-SUMM]{V-SUM \Metamodel}
    \acro{CPS}{Cyber-Physical System}
    \acro{CPR}{Consistency Preservation Rule}
    \acro{DSL}{Domain-Specific Language}
    \acro{EMF}{Eclipse Modeling Framework}
    \acro{OSM}{Orthographic Software Modeling}
    \acro{SUM}{Single Underlying Model}
    \acro{SUMM}{Single Underlying \Metamodel}
    \acro{EMF}{Eclipse Modeling Framework}
    \acro{2PL}{Two-Phase Locking}
    \acro{ACID}{Atomicity, Consistency, Isolation, Durability}
    \acro{SPL}{Software Product Line}
    \acro{SPLE}{Software Product Line Engineering}
    \acro{JSON}{JavaScript Object Notation}
    \acro{HTTP}{HyperText Transfer Protocol}
    \acro{VCS}{Version Control System}
    \acro{SQL}{Structured Query Language}
    \acro{OWL}{Web Ontology Language}
\end{acronym}

%% file: sections/01_introduction.tex
\section{Introduction}
\label{sec:introduction}

In Model-Driven Development (MDD), a single system is often described by multiple heterogeneous models that capture distinct concerns such as structure, behavior, or domain-specific properties. Because these models overlap conceptually, changes to one model require corresponding adjustments in others to maintain inter-model consistency~\cite{atkinson2003model,paige2016evolving}.

Multi-model consistency management is studied under various paradigms, including view-based development~\cite{klare2021enabling}, triple graph grammars~\cite{hermann2012concurrent}, and bidirectional transformations~\cite{stevens2020maintaining}. These approaches rely on Consistency Preservation Rules (CPRs) that automatically propagate changes between related models in order to re-establish consistency after modifications. Many multi-model engineering environments realize this by maintaining a logically integrated representation of the system and applying deterministic propagation rules that update dependent models after user-authored changes, namely \emph{original} changes. For example, the \vitruv framework~\cite{klare2021enabling} maintains consistency through a Virtual Single Underlying Model (\vsum), where CPRs automatically propagate modifications across model boundaries.

While such approaches enable consistency preservation during \emph{sequential} model evolution, they face fundamental challenges when combined with model versioning \cite{brosch2012introduction} and collaborative development. In practice, engineers routinely evolve systems concurrently on separate branches in order to explore alternative design solutions or work on different concerns in parallel. Integrating these parallel evolutions requires merging model changes originating from different branches. In multi-model settings, however, a merge must not only reconcile concurrent edits but also preserve inter-model consistency across all affected artifacts.

Traditional model versioning and merge tools do not account for the presence of cross-model consistency rules. State-based approaches such as EMF~Compare~\cite{emfcompare} operate on model snapshots and therefore cannot distinguish between \emph{original} user-authored changes and \emph{consequential} changes that were automatically produced by CPRs. Operation-based systems such as EMFStore~\cite{koegel2010emfstore} record edit histories but typically address single-model evolution without awareness of cross-model change propagation. As a result, both classes of techniques frequently report conflicts that arise solely from automatically generated state updates. Conflicts reported to a developer are not free: in an empirical study of code merges, Brindescu et al.~\cite{brindescu2020mergeconflicts} find that most conflicts require writing new code to resolve, interrupting the developer's current task, and that lines involved in a merge conflict are subsequently more defect-prone. That study concerns textual merges of source code, and we are not aware of comparable evidence for multi-model settings, so it should be read as motivation rather than as a quantification of the effort our approach saves; establishing the latter would require a user study, which we leave to future work. For instance, when two independent branches introduce changes that trigger consistency-preserving transformations affecting the same model element, snapshot- or history-based merges detect syntactic conflicts even though the propagated state would be regenerated deterministically during a consistency-preserving integration \cite{sharbaf2023conflict, ghiotto2020nature}.

This observation highlights a semantic gap between model versioning and multi-model consistency management: while existing merge techniques assume that all conflicting edits reflect deliberate user intent, automatically propagated changes merely reflect the application of consistency rules and should therefore not participate in merge decisions in the same way as original edits. Shen et al. propose a two‑stage approach for automated conflict resolution that incorporates similar historical fixes mined from open‑source repositories, thereby attempting to capture developer expertise and implicitly embedded knowledge about changes during merge resolution \cite{shen2023gitmerge}.

In this paper, we present a \emph{replay-based three-way merge} for multi-model consistency systems whose change metamodel distinguishes original from consequential changes. The approach is applicable to any system that (i)~maintains consistency via a deterministic application function and (ii)~records change provenance, i.e., whether each change was authored by the user or produced by a consistency rule. Rather than merging model states, our approach records the original changes of each commit and interleaves them from both branches through the consistency-preservation mechanism, using a dependency graph to find an ordering that preserves both branches' original intent. Consequential changes are thereby regenerated by the same rules that produced them originally, and only genuine conflicts between original changes require human intervention.

We formalize this merge operator, show that successful merges preserve global consistency and both branches' original intent, and identify two provenance-based conflict types that determine the merge ordering. A prototype implementation extends \vitruv with semantic changelog capture, UUID-based cross-branch conflict detection, and a per-commit replay engine with dependency-graph-based interleaving. We evaluate on three three-model case studies from different engineering domains, two automotive and one software-engineering, and compare our approach against EMF~Compare.

To discuss the contributions,
\Cref{sec:foundations} introduces the foundations of multi-model consistency and model versioning.
\Cref{sec:example} presents the motivating example.
\Cref{sec:formalization} formalizes the merge operator and its properties.
\Cref{sec:implementation} describes the prototype.
\Cref{sec:evaluation} reports the evaluation.
\Cref{sec:relatedwork} discusses related work, and
\cref{sec:conclusion} concludes the article.

%% file: sections/02_foundation.tex
\section{Foundations}
\label{sec:foundations}

This section introduces the concepts that our work builds upon: multi-model consistency management, consistency preservation rules, and model versioning with branching and merging.

\subsection{Multi-Model Consistency and View-Based Development}
\label{sec:foundations:vsum}

In model-driven development, a system is described by multiple models conforming to different metamodels. These models may conceptually overlap (e.g., structural and deployment models referencing the same components). A \emph{Consistency Relation}~$(CR)$ defines which combinations of model states are considered consistent~\cite{klare2022building,PascualBeckertUlbrichKirstenPfeifer2024}.

Our merge approach applies to multi-model consistency management mechanisms that (i)~maintain consistency via a deterministic application function that produces a consistent state from original changes, and (ii)~distinguish \emph{original changes} authored by the user from \emph{consequential changes} produced automatically by consistency rules. Weber et al.~\cite{weber2025how} survey change metamodels in model-driven engineering and identify consistency preservation as one purpose of change modeling; none of the surveyed metamodels records provenance at this original-vs.-consequential level. That survey is scoped to change metamodels rather than to merge implementations, so it bounds what is \emph{representable} rather than what any particular tool does. We therefore also examined multi-model infrastructures directly, DesignSpace with its reactive links, and Openflexo, and none records provenance at this level, though for different reasons in each case (\cref{sec:relatedwork}). A systematic review of 32~collaborative MDSE technologies~\cite{choudhury2025landscape} moreover finds only three that offer a diff/merge engine at all, and does not include change provenance among the capabilities by which such technologies are classified. We use \vitruv{} as the running instantiation throughout this paper because it records the distinction natively, not because we claim it is the only framework that could: requirement~(ii) is a property of the change representation, and \cref{sec:formalization:scope} states what a framework must provide to obtain the merge described here.

\begin{figure}[t]
    \centering
    \includegraphics[width=\columnwidth]{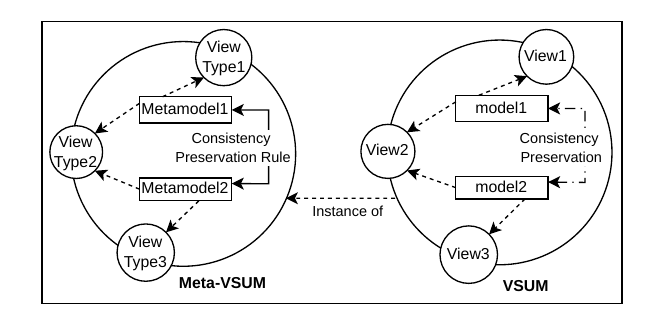}
    \vspace{-0.5cm}
    \caption{\vsum Concept in \vitruv}
    \label{fig:vsum}
    \vspace{-0.5cm}
\end{figure}

The \vitruv framework~\cite{klare2021enabling} maintains inter-model consistency through a \emph{Virtual Single Underlying Model} (\vsum) that integrates models, correspondences, and \emph{consistency preservation rules} (CPRs). Users interact with the system through \emph{views}; upon committing modifications, changes are propagated automatically to restore consistency (Figure~\ref{fig:vsum}).

\subsection{Consistency Preservation Rules (CPRs)}
\label{sec:foundations:cprs}

Consistency is maintained by CPRs, specified in \vitruv as \emph{Reactions}~\cite{klare2021enabling}. A Reaction is triggered by specific changes in one model and propagates updates to related models (e.g., creating a safety entry for a newly introduced component).

When a user applies changes~$\delta_1, \dots, \delta_n$ to a consistent state~$S$, \vitruv applies them and executes all applicable Reactions, potentially producing further changes~$\delta_{n+1}, \dots, \delta_{n+m}$. The full set of changes produced during a commit is written~$\tau$. Distinguishing between \emph{original} (user-authored) and \emph{consequential} (Reaction-produced) changes is central to our merge approach (see \cref{sec:formalization}).

CPRs can be combined into \emph{transformation networks} whose sequential application preserves global consistency under well-defined compatibility conditions~\cite[Chap.~4,5]{klare2022building}. We rely on this property as a foundational assumption.

\subsection{Model Versioning and Branching}
\label{sec:foundations:versioning}

Model versioning tracks the evolution of models over time~\cite{brosch2012introduction,mens2002survey,altmanninger2009survey}. \emph{State-based} approaches derive edit operations by comparing model snapshots (e.g., EMF~Compare~\cite{emfcompare}), while \emph{operation-based} approaches record explicit edit operations, preserving change order and intent~\cite{koegel2010emfstore,yohannisChangebasedModelDifferencing2020}.

Model differencing forms the basis for conflict detection and merging in version control systems~\cite{brosch2012introduction}. \emph{Branching} enables parallel development, while a \emph{three-way merge} reconciles diverged branches relative to a common ancestor; overlapping edits to shared elements constitute conflicts~\cite{mens2002survey}. Operation-based representations, such as EMF-CBP, further enable change-level conflict detection by capturing model evolution explicitly~\cite{yohannisChangebasedModelDifferencing2020}.

Single-model tools including EMF~Compare~\cite{emfcompare}, EMF~Diff/Merge, and AMOR~\cite{brosch2012introduction} support three-way merging with varying conflict detection capabilities.
However, they do not account for inter-model consistency relationships.
Automatically generated consequential changes produced by consistency rules are indistinguishable from original user edits, which may lead to false conflicts on elements whose state would otherwise be deterministically regenerated.

Other model versioning tools, such as the ones identified in a recent survey by~\citeauthor{kasaei2026benchmarv}\cite{kasaei2026benchmarv} also support branching workflows, though they either
do not make branching explicit~\cite{koegel2010emfstore}, only work on specific model types~\cite{DBLP:conf/staf/PietronFT21}, or do not work with multiple, separate models that are kept consistent~\cite{kellyCollaborativeModellingVersion2018}.

%% file: sections/03_illustratingexample.tex
\section{Running Example}
\label{sec:example}

We illustrate and later evaluate our approach with an automotive brake system engineering scenario, based on a case study by \citeauthor{hagel2025brakecasestudy}~\cite{hagel2025brakecasestudy}.
Their case study provides metamodels for cross-domain brake system engineering involving product specification, CAD, and simulation models.
We reuse and extend their brake specification and CAD metamodels, and add a safety assessment metamodel with \vitruv Reactions that maintain consistency across all three models.

\subsection{Automotive Brake System Meta-\vsum}
Figure~\ref{fig:vsum-consequential-overlap} shows the Meta-\vsum for the Automotive Brake System. The models are as follows:

\begin{figure}[t]
\centering
\input{figures/vsum_consequential_overlap}
\caption{Relevant elements of the Meta-\vsum for the Automotive Brake System}
\label{fig:vsum-consequential-overlap}
\vspace{-0.5cm}
\end{figure}
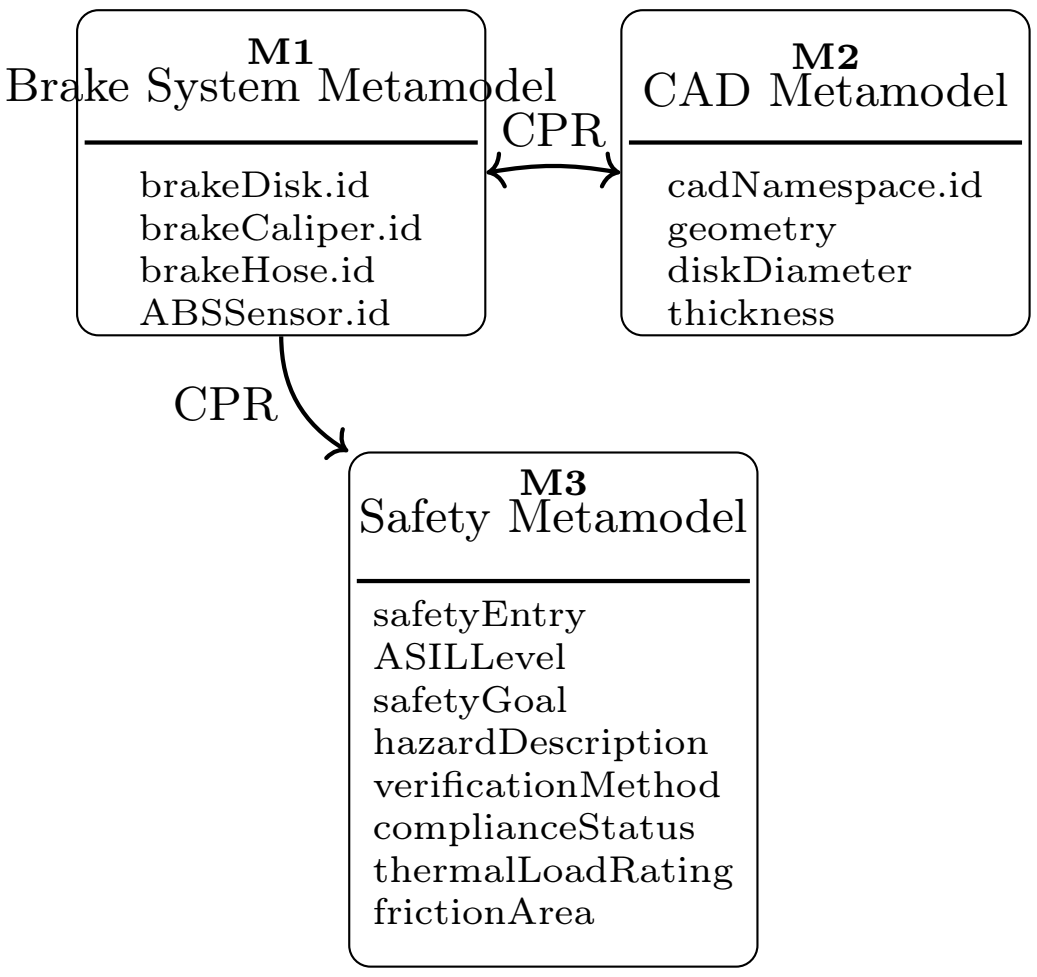

\paragraph{$M_1$: Brake System Model.}
Based on the brake specification metamodel of Hagel et al.~\cite{hagel2025brakecasestudy}, the primary engineering model describes the brake system's components: \texttt{BrakeDisk} (with attributes such as \texttt{diameterInMM} and \texttt{brakeDiskThicknessInMM}), \texttt{BrakeCaliper}, \texttt{BrakePad} (with \texttt{heightInMM}, \texttt{widthInMM}), \texttt{BrakeHose}, and \texttt{ABS\-Sensor}. Each component carries an \texttt{id} attribute (a domain-level label, distinct from the framework-level element identifiers used for conflict detection in \cref{sec:formalization}).

\paragraph{$M_2$: CAD Model.}
Derived from the CAD metamodel of Hagel et al.~\cite{hagel2025brakecasestudy}, the CAD representation mirrors the brake system's structure as \texttt{Namespaces}, each containing typed \texttt{Parameters} (numeric, string, boolean). A \texttt{Namespace} corresponds to a component; its \texttt{Parameters} correspond to the component's attributes.

\paragraph{$M_3$: Safety Model.}
We extend the original case study with a safety assessment model that holds a \texttt{SafetyEntry} per component with derived quantities \texttt{thermalLoadRating} (computed from disk diameter and thickness) and \texttt{frictionArea} (computed from brake pad dimensions), plus manually completed fields such as ASIL level and safety goal. Reactions from~$M_1$ to~$M_3$ automatically create safety entries for new components and maintain derived metrics when engineering parameters change.

\subsection{Consistency Rules}

Three sets of \vitruv Reactions maintain consistency:
\begin{itemize}
    \item $M_1 \to M_2$: Creating a brake component creates a CAD namespace; changing an attribute updates the corresponding CAD parameter; renaming a component's \texttt{id} updates the namespace's \texttt{id}.
    \item $M_2 \to M_1$: Renaming a namespace's \texttt{id} propagates back to the component's \texttt{id} (bidirectional).
    \item $M_1 \to M_3$: Creating a component creates a safety entry; changing the disk's diameter or thickness recomputes \texttt{thermalLoadRating}; changing pad dimensions recomputes \texttt{frictionArea}.
\end{itemize}

\noindent For instance, when an engineer sets \texttt{BrakeDisk.diameterInMM} to 320, Reactions automatically update the CAD parameter ``Diameter'' to 320 and recompute the safety entry's \texttt{thermalLoadRating}.

\subsection{Branching and Merge Scenarios}

Suppose two engineers work on separate branches, starting from a common state that contains a brake disk with \texttt{diameterInMM}~=~300.

\paragraph{Scenario 1: Clean merge.}
Engineer~A changes the disk diameter from 300 to 320 in~$m_1$. Reactions derive updates to~$m_2$ (CAD Diameter~=~320) and~$m_3$ (\texttt{thermalLoadRating} recomputed). Engineer~B adds an \texttt{ABSSensor} component, which triggers creation of a new CAD namespace and a new safety entry. The changes affect disjoint elements; the merge succeeds without conflicts.

\paragraph{Scenario 2: Direct conflict (original vs.\ original).}
Both engineers change \texttt{BrakeDisk.diameterInMM}: A sets it to 320, B sets it to 350. Both changes are original and target the same element and feature. This is a \emph{direct conflict} requiring the user to choose a value.

A state-based tool such as EMF~Compare detects this conflict in~$m_1$, but additionally reports conflicts in~$m_2$ (both branches produced different consequential CAD Diameter values) and~$m_3$ (both produced different consequential \texttt{thermalLoadRating} values), yielding three conflicts total. Our approach detects only the single root conflict in~$m_1$, because the conflicts on consequential elements in~$m_2$ and~$m_3$ would be resolved automatically once the user resolves the root cause.

\paragraph{Scenario 3: Consequential overlap (consequential vs.\ original).}
Engineer~A renames \texttt{BrakeDisk.id} from \texttt{disk1} to \texttt{engineeringDisk} in~$m_1$. The $M_1 \to M_2$ Reaction propagates this as a namespace rename. Meanwhile, Engineer~B directly sets the CAD namespace \texttt{id} to \texttt{cadCustomId} in~$m_2$ (an original change). When A's changes are replayed onto B's branch, the consequential namespace rename overwrites B's explicit choice. Our approach detects this as a \emph{consequential overlap}: a consequential change from replay overlaps with original intent on the target branch.
EMF~Compare, unable to distinguish original from consequential, reports five hard conflicts across all three models.

%% file: figures/vsum_consequential_overlap.tex

\begin{tikzpicture}[
font=\footnotesize,
class/.style={
    draw,
    rounded corners,
    minimum width=2.7cm,
    inner sep=2pt
},
divider/.style={thick},
cpr/.style={->, thick}
]

\node[
draw,
rounded corners,
minimum width=2.7cm,
minimum height=2.15cm
] (M1) at (-1.8,1.35) {};

\node[align=center] at (-1.8,2.0) {
\textbf{M1}\\[-1pt]
{\scriptsize Brake System Metamodel}
};

\draw[divider] (-3.1,1.55)--(-0.5,1.55);

\node[align=left] at (-1.8,0.85) {
brakeDisk.id\\
brakeCaliper.id\\
brakeHose.id\\
ABSSensor.id
};

\node[
draw,
rounded corners,
minimum width=2.7cm,
minimum height=2.15cm
] (M2) at (1.8,1.35) {};

\node[align=center] at (1.8,2.0) {
\textbf{M2}\\[-1pt]
{\scriptsize CAD Metamodel}
};

\draw[divider] (0.5,1.55)--(3.1,1.55);

\node[align=left] at (1.8,0.85) {
cadNamespace.id\\
geometry\\
diskDiameter\\
thickness
};

\node[
draw,
rounded corners,
minimum width=2.7cm,
minimum height=3.4cm
] (M3) at (0,-2.2) {};

\node[align=center] at (0,-0.85) {
\textbf{M3}\\[-1pt]
{\scriptsize Safety Metamodel}
};

\draw[divider] (-1.3,-1.35)--(1.3,-1.35);

\node[align=left] at (0,-2.55) {
safetyEntry\\
ASILLevel\\
safetyGoal\\
hazardDescription\\
verificationMethod\\
complianceStatus\\
thermalLoadRating\\
frictionArea
};


\draw[<->, thick]
(M1.east) to[bend left=10]
node[midway, above] {\scriptsize CPR}
(M2.west);

\draw[cpr]
(M1.south) to[bend right=30]
node[midway, left] {\scriptsize CPR}
(M3.north west);

\end{tikzpicture}


%% file: sections/04_formalization.tex
\section{Replay-Based Merge}
\label{sec:formalization}

This section formalizes a replay-based three-way merge for proven\-ance-aware multi-model consistency systems. The formalization is independent of \vitruv and applies to any system satisfying the prerequisites in \cref{sec:foundations:vsum}. We define the merge operator, identify two provenance-based conflict types, introduce guarded replay and a dependency-graph-based interleaving algorithm, and state correctness properties.

\Cref{tab:notation} collects the notation used throughout. We distinguish three levels consistently: $\delta$ is a single atomic change, $\tau$ a sequence of them recorded by one commit, and $H$ a sequence of commits forming a branch history.

\begin{table}[t]
\centering
\caption{Notation used in \cref{sec:formalization}.}
\label{tab:notation}
\small
\begin{tabular}{ll}
\hline
\textbf{Symbol} & \textbf{Meaning} \\
\hline
$M_i$, $m_i$ & metamodel $i$ and a conforming model instance \\
$\mathcal{I}$ & space of global model states \\
$S$, $S_0$ & a global state; the common ancestor state \\
$CR$ & consistency relation, $CR \subseteq \mathcal{I}$ \\
$\mathrm{Cons}(S)$ & $S$ is globally consistent, i.e.\ $S \in CR$ \\
\hline
$o$ & an atomic operation (\cref{tab:operations}) \\
$\delta = (o, \rho)$ & an atomic change: operation plus provenance \\
$\rho \in \{\mathsf{orig}, \mathsf{cons}\}$ & provenance label of a change \\
$\delta^o$, $\delta^c$ & shorthand for a change with $\rho=\mathsf{orig}$ / $\mathsf{cons}$ \\
$\tau$ & a commit: the change sequence of one version-control commit \\
$\tau^o$, $\tau^c$ & its original / consequential projections (\cref{def:projection}) \\
$\Delta_O$, $\Delta_C$ & spaces of original / consequential change sequences \\
$H$, $H_A$, $H_B$ & a history of commits; the histories of branches $A$, $B$ \\
\hline
$APP$ & consistency-preserving application function \\
$APP^*$ & iterated application of a history \\
$\pi_S$, $\pi_C$ & projections of $APP$ to state / consequential changes \\
$\gamma$ & applicability predicate (guard) of a commit \\
\hline
$\mathcal{F}$ & set of all (element, feature) pairs \\
$\mathit{fp}(\cdot)$ & footprint: the pairs in $\mathcal{F}$ written \\
$\mathit{fp}_O$, $\mathit{fp}_C$ & original / consequential footprint of a commit \\
$\#$ & footprint conflict (same pair, different values) \\
$G$, $\sigma$ & footprint dependency graph; a topological ordering of it \\
\hline
\end{tabular}
\vspace{-0.3cm}
\end{table}

\subsection{Model Space and Consistency}

Let $\vec{M} = (M_1, \dots, M_n)$ be a tuple of metamodels and
\[
S = (m_1, \dots, m_n) \in \mathcal{I} := I_{M_1} \times \cdots \times I_{M_n}
\]
a global model state, where $I_{M_i}$ denotes the instances of $M_i$. We use $M_i$ for metamodels and $m_i$ for the corresponding model instances throughout.
A \emph{consistency relation} $CR \subseteq \mathcal{I}$ defines the globally consistent states, i.e., $\mathrm{Cons}(S) \Leftrightarrow S \in CR$.
This follows the tuple-based view used in \vitruv~\cite{klare2021enabling,klare2022building} and related formalizations~\cite{PascualBeckertUlbrichKirstenPfeifer2024}.
The definitions in Table~\ref{tab:notation} are framework-independent and require only concrete definitions of $CR$ (Consistency Relation) and $APP$ (Application Function).

\paragraph{Example.}
In the running example (\cref{sec:example}), $\vec{M} = (M_1, M_2, M_3)$ correspond to the Brakesystem, CAD, and Safety metamodels. A consistent state pairs, e.g., a \texttt{BrakeDisk} with \texttt{diameterInMM}~=~300 in~$m_1$ with a CAD namespace whose Diameter equals 300 in~$m_2$, and the corresponding derived \texttt{thermalLoadRating} in~$m_3$.

\subsection{Atomic Operations, Changes, and Provenance}
\label{sec:formalization:changes}

Figure~\ref{fig:branchmerge} illustrates branching with sequences of changes~$\delta$ of two kinds: \emph{original} changes~$\delta^o$ authored by the user and \emph{consequential} changes~$\delta^c$ produced automatically by consistency rules. We first fix what a change \emph{is}, because both the footprint (\cref{sec:formalization:footprints}) and the conflict classification (\cref{sec:formalization:conflicts}) are derived from it.

\paragraph{Atomic operations.}
Let $E$ be the set of model elements, each carrying a stable identity (\cref{sec:impl:uuid}), and let $\mathit{feat}(e)$ denote the features of~$e$. We take the change metamodel to provide the five atomic operations in \cref{tab:operations}, which correspond to the elementary edit operations of delta-based model repositories~\cite{brosch2012introduction,taentzer2014fundamental} and, in our prototype, to \vitruv's \texttt{EChange} hierarchy. Composite edits are sequences of these; we do not assume any coarser notion.

\begin{table}[t]
\centering
\caption{Atomic operations and the footprint each induces. $\mathit{cont}^*(e)$ denotes $e$ together with its transitively contained elements.}
\label{tab:operations}
\small
\begin{tabular}{lp{3.1cm}p{3.3cm}}
\hline
\textbf{Operation} & \textbf{Effect} & \textbf{Footprint $\mathit{fp}$} \\
\hline
$\mathsf{set}(e,f,v)$ & set single-valued feature $f$ of $e$ to $v$ & $\{(e,f)\}$ \\
$\mathsf{ins}(e,f,x)$ & insert $x$ into multi-valued feature $f$ of $e$ & $\{(e,f,x)\}$ \\
$\mathsf{rem}(e,f,x)$ & remove $x$ from multi-valued feature $f$ of $e$ & $\{(e,f,x)\}$ \\
$\mathsf{new}(e,p,f)$ & create $e$ and place it in feature $f$ of parent $p$ & $\{(p,f,e)\}$ \\
$\mathsf{del}(e)$ & delete $e$ and everything it contains & $\{(p,f,e)\} \cup \{(e',f') \mid e' \in \mathit{cont}^*(e),\; f' \in \mathit{feat}(e')\}$ \\
\hline
\end{tabular}
\vspace{-0.3cm}
\end{table}

\paragraph{Provenance is a label, not a kind.}
Original and consequential changes are built from the \emph{same} operations; they differ only in who produced them. We therefore define a change as a pair
\[
\delta = (o, \rho), \qquad \rho \in \{\mathsf{orig}, \mathsf{cons}\},
\]
where $o$ is an atomic operation and $\rho$ records whether $o$ was authored by an engineer or emitted by a consistency rule. We write $\delta^o$ and $\delta^c$ for changes labelled $\mathsf{orig}$ and $\mathsf{cons}$ respectively. Nothing distinguishes the two beyond the label: a consequential $\mathsf{set}$ and an original $\mathsf{set}$ have the same effect on the state and the same footprint. This matters for the scope of the approach, the merge operator is not restricted to some special class of derived change (\cref{sec:formalization:scope}).

A \emph{commit}~$\tau$ is the sequence of changes recorded together by one version-control commit.

\begin{definition}[Provenance projection]
\label{def:projection}
For a commit $\tau = (\delta_1,\dots,\delta_n)$, the \emph{original} and \emph{consequential} projections are the subsequences
\[
\tau^o = (\delta_i)_{\rho_i = \mathsf{orig}}, \qquad
\tau^c = (\delta_i)_{\rho_i = \mathsf{cons}},
\]
so that $\tau$ is recovered by interleaving them in their recorded order.
\end{definition}

Because engineers author changes and rules react to them, each original change is followed by the consequential changes it triggers, and $\tau = \tau^o \cdot \tau^c$ up to that ordering. $\Delta_O$ and $\Delta_C$ denote the sets of sequences in the image of the respective projection; they are views of one history rather than two disjoint sources of change.

\paragraph{Application.}
\vitruv provides a consistency-preserving application function
\[
APP : \mathcal{I} \times \Delta_O \rightharpoonup \mathcal{I} \times \Delta_C,
\]
where $APP(S,\tau^o)=(S',\tau^c)$ applies $\tau^o$ in state~$S$, executes all triggered Reactions, yields the consistent state~$S'$, and produces consequential changes~$\tau^c$.

\begin{assumption}[Consistency Preservation and Determinism]
If $\mathrm{Cons}(S)$ and $APP(S,\tau^o)$ is defined, then $\mathrm{Cons}(S')$, and the result of $APP(S,\tau^o)$ is uniquely determined.
\end{assumption}

Determinism is state-dependent: replaying the same original changes in a different state may produce different consequential changes, which is the source of consequential conflicts (\cref{sec:formalization:conflicts}). In the running example, setting \texttt{BrakeDisk.diameterInMM} to 320 triggers updates in $m_2$ and recomputation in $m_3$; these consequential changes are regenerated on replay and need not be stored explicitly.

\input{figures/branch-merge-figure}

\subsection{Sequential Evolution and Histories}

Let $H = (\tau^o_1, \dots, \tau^o_k)$ be a history of original commits.
Iterative application of $H$ on a state $S_0$ is defined as
\begin{align*}
APP^*(S_0, ()) &= S_0, \\
APP^*(S_0, (\tau^o_1, \dots, \tau^o_k)) &=
\pi_S(APP(APP^*(S_0, (\tau^o_1, \dots, \tau^o_{k-1})), \tau^o_k)).
\end{align*}

We denote the resulting state by $S_H = APP^*(S_0,H)$.

\subsection{Branches and Divergence}

Let $S_0$ be a consistent common ancestor. For two original-commit
histories in branches $A$ and $B$
\[
H_A=(\tau^o_{A1},\dots,\tau^o_{Ap}), \quad
H_B=(\tau^o_{B1},\dots,\tau^o_{Bq}),
\]
the corresponding branch heads are
\[
S_A=APP^*(S_0,H_A), \quad
S_B=APP^*(S_0,H_B).
\]

\paragraph{Example.}
In Scenario~2 (\cref{sec:example}), $S_0$ contains a brake disk with
\texttt{diameterInMM}=300. Engineer~A sets the value to 320
($H_A=(\tau^o_{A1})$), while Engineer~B sets it to 350
($H_B=(\tau^o_{B1})$). The resulting heads $S_A$ and $S_B$ diverge
both on this attribute and on consequential downstream values in
$m_2$ and $m_3$.

\subsection{Conflict Footprints}
\label{sec:formalization:footprints}

Let $\mathcal{F}$ be the set of all (element, feature) pairs across all model instances, where elements are identified by stable UUIDs (\cref{sec:impl:uuid}); for multi-valued features we additionally qualify the pair by the inserted or removed element, writing $(e,f,x)$, so that independent insertions remain distinguishable.
The footprint of an atomic change is \emph{derived} from its operation, as given in the third column of \cref{tab:operations}; it is not an independent primitive. The footprint of a change $\delta=(o,\rho)$ is that of its operation, $\mathit{fp}(\delta) := \mathit{fp}(o)$, independently of the provenance label. Footprints lift to sequences by union: for a commit $\tau=(\delta_1,\dots,\delta_n)$, $\mathit{fp}(\tau) := \bigcup_i \mathit{fp}(\delta_i)$, and we write $\mathit{fp}_O(\tau) := \mathit{fp}(\tau^o)$ and $\mathit{fp}_C(\tau) := \mathit{fp}(\tau^c)$ for the footprints of the two provenance projections.

Two changes conflict, written
$\mathit{fp}(\delta_1)\#\mathit{fp}(\delta_2)$, if their footprints share a pair that they write with different values; deletion is treated as writing a distinguished absent value, so a delete and a modify of the same pair conflict.
For multi-valued features, insertions of distinct elements produce disjoint footprints and are therefore non-conflicting under this definition (cf.\ assumption~(iii) in \cref{sec:formalization:scope}).

\paragraph{Creation and deletion.}
Two entries of \cref{tab:operations} deserve comment, because they are what makes the definition total rather than restricted to attribute updates.

A creation $\mathsf{new}(e,p,f)$ writes the \emph{parent's} containment feature, not the new element: before the operation, $e$ does not exist and cannot be the subject of a write. Two branches independently creating elements therefore touch $(p,f,e_A)$ and $(p,f,e_B)$ with distinct fresh identities, which do not conflict, the formal counterpart of the observation that adding an \texttt{ABSSensor} on one branch never blocks the other.

A deletion $\mathsf{del}(e)$ writes both the containment slot that held $e$ and \emph{every} feature of $e$ and of its transitively contained elements. The cascade is essential: without it, deleting a container on one branch while the other modifies a nested child would produce disjoint footprints and the conflict would be missed. With it, the delete/modify case requires no special rule, it is an ordinary footprint overlap, and \cref{sec:formalization:conflicts} classifies it as a direct conflict whenever both sides are original. This matches the implementation, which detects deletions by element identity rather than per feature and records the identities of cascade-deleted children at capture time (\cref{sec:impl:changelog}).

Note the distinction from \cref{def:guard-failure}: a delete/modify pair in which \emph{both} changes are original is a direct conflict, decided before replay. A modification that fails because a \emph{consequential} change removed its target is not a conflict of intent but a guard failure, resolved by re-ordering where possible.

\paragraph{Example.}
Engineer~A's commit $\tau^o_{A1}$ (set
\texttt{diameterInMM}=320) has footprint
$\mathit{fp}(\tau^o_{A1})=\{(\mathit{uuid}_{\mathrm{disk}},
\texttt{diameterInMM})\}$.
Engineer~B's $\tau^o_{B1}$ (set \texttt{diameterInMM}=350)
targets the same element-feature pair with a different value,
so $\mathit{fp}(\tau^o_{A1})\#\mathit{fp}(\tau^o_{B1})$ holds (Scenario~2).
In Scenario~1, adding a new \texttt{ABSSensor} yields a fresh
UUID and does not conflict.

\subsection{Guarded Replay and Applicability}
\label{sec:formalization:guards}

Consequential changes are regenerated during replay. Thus, an original commit valid in its source context may no longer be applicable on the target branch.

\begin{definition}[Guarded Original Commit]
A guarded commit $\hat{\tau}=(\tau,\gamma)$ consists of an original change sequence $\tau \in \Delta_O$ and an applicability predicate $\gamma:\mathcal{I}\to\{\mathit{true},\mathit{false}\}$ indicating whether the elements and features referenced by~$\tau$ exist in the given state.
\end{definition}

Replaying a sequence of guarded commits from an initial state $\widetilde{S}_0$ yields
\[
\widetilde{S}_i=
\begin{cases}
\pi_S(APP(\widetilde{S}_{i-1},\tau^o_i)) & \text{if }\gamma_i(\widetilde{S}_{i-1}),\\
\bot & \text{otherwise.}
\end{cases}
\]

\begin{definition}[Guard Failure]
\label{def:guard-failure}
A guard failure occurs when a commit's applicability predicate holds in its source pre-state but not in the replay state.
\end{definition}
\paragraph{Example.} Modifying an element that is absent from the replay state causes a guard failure. Note that this is the \emph{residual} case only: if the element was deleted by an original change on the other branch, the footprints overlap by \cref{tab:operations} and the pair is already reported as a direct conflict before replay begins (\cref{sec:formalization:footprints}). A guard failure arises when no engineer authored the deletion, the element was removed by a \emph{consequential} change under the current ordering. That is not a conflict between change intentions but an operational precondition failure, and the interleaving algorithm (\cref{sec:formalization:interleaving}) attempts to resolve it by finding an alternative ordering.

\subsection{Conflicts}
\label{sec:formalization:conflicts}

The merge algorithm distinguishes two types of conflicts:

\begin{description}

\item[Direct conflict (original vs.\ original):]
    Both branches write the same element-feature pair with different
    values through \emph{original} changes, i.e., $\mathit{fp}_O(a_i) \# \mathit{fp}_O(b_j)$.
    Direct conflicts are detected statically from the changelog DTOs
    before replay and require human resolution.

\item[Consequential conflict:]
    The dependency graph (\cref{sec:formalization:interleaving})
    contains a cycle. A cycle arises when consequential footprints
    create ordering constraints in both directions: commit~$a_i$'s
    Reactions write an element-feature pair that~$b_j$ modifies
    originally, and~$b_j$'s Reactions write a pair that~$a_i$ modifies
    originally. No interleaving can preserve both branches' original
    intent; human resolution is required.
    This is analogous to a serialization conflict in database
    concurrency control~\cite{bernstein1987concurrency}, where a cycle
    in the serialization graph prevents a serializable schedule.
\end{description}

\noindent\emph{Counting unit.} Both conflict types are counted per element-feature pair. A direct conflict is one pair modified by both branches with different values. A cycle in the dependency graph typically involves multiple overlapping element-feature pairs (one per direction of the cycle); each is counted as a separate blocking conflict, since each requires independent resolution.

If the graph is acyclic, consequential overlaps are resolved automatically by the ordering; a cycle constitutes one or more consequential conflicts. Guard failures are a separate operational concern (\cref{def:guard-failure}).

\paragraph{What the cycle criterion buys.}
It is worth stating plainly what this classification provides that a state-based comparison cannot, since both ultimately hand the engineer a set of locations to inspect.

A state-based tool compares $S_A$ and $S_B$ against $S_0$ and reports every location where they differ. It has no access to provenance, so it cannot tell an element the engineer wrote from one a Reaction produced, and it therefore cannot form $G$ at all: the edge condition $\mathit{fp}_C(a_i)\,\#\,\mathit{fp}_O(b_j)$ is not expressible over states. Lacking that, it must treat every difference as a candidate conflict, including differences on elements that a consistency-preserving replay would simply regenerate. This is not a deficiency of any particular implementation but a consequence of the input: the information that distinguishes the two cases is discarded when a history is reduced to the state it produced.

The provenance-aware criterion answers a different and stricter question. An acyclic $G$ is a \emph{constructive} verdict: it exhibits an ordering under which every original change from both branches is the last write to each pair it targets, so the merge is not merely plausible but justified by \cref{prop:intent}. A cycle is an equally precise negative verdict: it certifies that \emph{no} interleaving preserves both branches' original intent, so the conflict is a genuine incompatibility of engineering decisions rather than an artifact of how the models were serialized. The unit of the report changes accordingly, from ``these locations differ'' to ``these two design decisions cannot both stand'', which is what makes the count in \cref{sec:evaluation} comparable in kind rather than merely smaller.

\paragraph{Example.}
Engineer~A renames
\texttt{BrakeDisk.id} in~$m_1$ and Engineer~B renames
\texttt{Namespace.id} in~$m_2$. These are different element-feature
pairs, so no direct conflict exists. However, A's Reaction propagates
the rename to \texttt{Namespace.id} (overlapping B's original change),
and B's Reaction propagates back to \texttt{BrakeDisk.id} (overlapping
A's original change). The dependency graph has edges
$a_1 \to b_1$ and $b_1 \to a_1$, forming a cycle.

\subsection{Merge via Interleaving}
\label{sec:formalization:interleaving}

The merge of branches $A$ and $B$ finds an \emph{interleaving} of their original commit histories that preserves both branches' original intent.

\begin{definition}[Interleavings of Histories]
An interleaving of $H_A=(\tau^o_{A1},\ldots,\tau^o_{Ap})$ and
$H_B=(\tau^o_{B1},\ldots,\tau^o_{Bq})$ is any permutation of
$H_A\cup H_B$ that preserves intra-branch ordering.
\end{definition}

Let $p=|H_A|$ and $q=|H_B|$. There are $\binom{p+q}{p}$ possible interleavings. Rather than enumerating them, we use a dependency graph to find a valid one directly.

\paragraph{Footprint dependency graph.}
We build a dependency graph $G=(V,E)$ inspired by serialization graphs in database concurrency control~\cite{bernstein1987concurrency}:
\begin{itemize}
    \item $V=\{a_1,\ldots,a_p,b_1,\ldots,b_q\}$, one node per commit. Nodes represent only original changes; consequential changes are regenerated during replay.
    \item Edges $a_i\to b_j$ exist when
    $\mathit{fp}_C(a_i)\#\mathit{fp}_O(b_j)$, and symmetrically for $b_j\to a_i$.
\end{itemize}
Here, $\mathit{fp}_C(a_i)$ denotes the \emph{consequential footprint} of commit~$a_i$: the set of element-feature pairs modified by Reactions when~$a_i$ is replayed.
An edge $a_i\to b_j$ encodes that $a_i$ must be replayed \emph{before}~$b_j$: this way, $a_i$'s consequential change is applied first, and $b_j$'s original change, applied later, overwrites it, preserving $b_j$'s original intent.

A cycle in $G$ indicates a consequential conflict (no ordering avoids all consequential overlaps);
otherwise, a topological sort (e.g., Kahn's algorithm~\cite{kahn1962topological})
yields a valid interleaving.

\input{figures/dependency-graph-figure}

\paragraph{Worked construction.}
\Cref{fig:depgraph} constructs the graph for the two scenarios that differ only in the direction of the Reactions, which is what makes one resolvable and the other not.

In both, branch~$A$'s commit $a_1$ renames \texttt{BrakeDisk.id} in~$m_1$ and branch~$B$'s commit $b_1$ renames the corresponding CAD \texttt{Namespace.id} in~$m_2$. Neither is a direct conflict: the two original footprints, $\mathit{fp}_O(a_1)=\{(\mathrm{disk},\texttt{id})\}$ and $\mathit{fp}_O(b_1)=\{(\mathrm{ns},\texttt{id})\}$, are disjoint, so no element-feature pair is written originally by both. A state-based comparison sees only the four differences that result and cannot go further.

The edges come from the \emph{consequential} footprints. Replaying $a_1$ fires the $M_1\!\to\!M_2$ Reaction, which writes $(\mathrm{ns},\texttt{id})$, exactly the pair $b_1$ writes originally. So $\mathit{fp}_C(a_1)\,\#\,\mathit{fp}_O(b_1)$ and we add $a_1\to b_1$, recording that $a_1$ must be replayed \emph{first} so that $b_1$'s original write is the last one to land on that pair (\cref{fig:depgraph}b). The topological order $\sigma=(a_1,b_1)$ discharges the constraint, and $b_1$'s original intent survives; this is \textsc{resolved\_overlap}.

Now suppose the $M_1\!\leftrightarrow\!M_2$ Reactions are bidirectional, as they are in the brake case study. Replaying $b_1$ then also fires a Reaction, writing $(\mathrm{disk},\texttt{id})$, which is what $a_1$ wrote originally. The symmetric condition $\mathit{fp}_C(b_1)\,\#\,\mathit{fp}_O(a_1)$ adds $b_1\to a_1$, and the graph now contains the cycle $a_1\to b_1\to a_1$ (\cref{fig:depgraph}c). The two constraints are jointly unsatisfiable: whichever commit is replayed last, the other's original write has been overwritten by a consequential one. This is \textsc{consequential}, and it is reported for human resolution.

\paragraph{Effect of re-ordering.}
The graph is what changes when commits are re-ordered, and it changes in one direction only: edges are added, never removed. Re-ordering does not alter $\mathit{fp}_O$, which is fixed by what the engineer wrote, but it does alter $\mathit{fp}_C$, because a Reaction fired in a different state may write different pairs. When replay reveals a consequential footprint larger than the estimate, or a guard fails, the corresponding edge is added and the graph is re-sorted. A previously acyclic graph can therefore become cyclic, which is why the conflict verdict is only final once the iteration has converged, and why termination rests on the monotonicity of the edge set rather than on a single pass.

\paragraph{Replay procedure.}
Given a topological ordering $\sigma=(\sigma_1,\ldots,\sigma_{p+q})$ of~$G$, replay proceeds from the common ancestor~$S_0$. For each commit $\sigma_k$:
\begin{enumerate}
    \item Check the guard $\gamma_k(\widetilde{S}_{k-1})$; if it fails, add a directed edge from the most recent other-branch commit~$\sigma_j$ ($j < k$) to~$\sigma_k$, rebuild, and re-sort. This is a heuristic: a guard failure means $\sigma_k$'s preconditions held on its source branch but not in the replay state, which was shaped by other-branch commits replayed before it. Adding the edge $\sigma_j \to \sigma_k$ forces the re-sort to place~$\sigma_k$ later, after more same-branch context that may restore the missing preconditions. The heuristic is not complete, a valid ordering may exist that this edge does not lead to, but it is sound: if no acyclic ordering remains, the algorithm correctly reports a consequential conflict.
    \item Replay via $APP(\widetilde{S}_{k-1},\sigma_k)=(\widetilde{S}_k,\tau^c_k)$; Reactions fire and produce consequential changes.
    \item Capture the actual consequential footprint $\mathit{fp}_C(\sigma_k)$ for comparison with the estimate after replay completes (see iterative refinement below).
\end{enumerate}
A replay is \emph{conflict-free} if no direct conflicts, consequential conflicts, or unresolved guard failures occur.

\paragraph{Iterative footprint refinement.}
Consequential footprints are state-dependent: replaying $\tau^o_i$ on a state $\widetilde{S}$ may produce different consequential changes than on the original branch. The initial footprint estimates (captured during branch work; see \cref{sec:impl:changelog}) serve as a starting point. After executing a topological ordering, the algorithm compares actual footprints against estimates. If new footprints or guard failures introduce new edges, the graph is rebuilt and re-sorted; if a cycle emerges, a consequential conflict is reported.

\emph{Termination.} Each iteration that does not succeed adds at least one new edge to~$G$. Edges are never removed, and the maximum number of inter-branch edges is $p \cdot q$. Hence the algorithm terminates in at most $p \cdot q + 1$ iterations: it either finds a valid ordering or detects a cycle.

\subsection{Formal Analysis of the Footprint Dependency Graph as a Serialization Graph}
\label{sec:formal_footprint_dependency_graph}

The footprint dependency graph $G$ (defined in Subsection~\ref{sec:formalization:interleaving}) can be formally analyzed as a serialization graph, a well-established construct from database theory and concurrent transaction scheduling \cite{bernstein1987concurrency}. This formalization provides a rigorous foundation for understanding validity of interleavings, detecting and resolving conflicts (cycles in $G$), and proving the correctness of the merge algorithm. Below, we map the concepts from Section~\ref{sec:formalization:interleaving} to their formal counterparts in serialization graph theory. In that mapping each commit plays the role of a \emph{transaction} in the database sense. We borrow only the scheduling machinery, serializability of an interleaving and cycle detection in the resulting graph, and not the ACID guarantees: a commit here is neither isolated nor atomic under replay, and a guard failure may leave it partially applied before re-ordering resolves it. The property we establish is consistency of the merged state, not atomicity, isolation, or durability.

\begin{itemize}
    \item \textbf{Graph Structure and Properties:}
    \begin{itemize}
        \item \textbf{Nodes ($V$):} Each node represents an original commit from either branch (e.g., $a_1, \dots, a_p$ from branch $A$ and $b_1, \dots, b_q$ from branch $B$).

        \item \textbf{Edges ($E$):} Edges in $G$ encode \emph{ordering constraints} between commits. These constraints are derived from the footprints of original and consequential changes:
        \begin{itemize}
            \item \textbf{Intra-branch edges:} These preserve the sequential order of commits within a branch (e.g., $a_i \to a_{i+1}$).

            \item \textbf{Inter-branch edges:} An edge $a_i \to b_j$ exists if and only if the \emph{consequential footprint} of commit $a_i$ overlaps with the \emph{original footprint} of commit $b_j$ (or vice versa). Formally, this is defined as:
            \[
            a_i \to b_j \iff \mathit{fp}_C(a_i) \# \mathit{fp}_O(b_j).
            \]
            This edge signifies that $a_i$'s consequential changes must be applied before $b_j$'s original changes to preserve intent.
        \end{itemize}
    \end{itemize}

    \item \textbf{Conflict Classification via Cycle Detection:}
    \begin{itemize}
        \item A cycle in $G$ indicates a \emph{consequential conflict}, where no valid interleaving can satisfy both branches' original intent without overwriting changes. This is analogous to a \emph{serialization conflict} in database theory \cite{bernstein1987concurrency}, where a cycle in the serialization graph prevents a \emph{serializable schedule}.
        \item In the context of the footprint dependency graph, a cycle arises when:
        \begin{itemize}
            \item The consequential footprint of commit $a_i$ overlaps with the original footprint of commit $b_j$.
            \item The consequential footprint of commit $b_j$ overlaps with the original footprint of commit $a_i$.
        \end{itemize}
        \item \textbf{Example:} In Scenario~3 (Section~\ref{sec:example}), Engineer $A$ renames $\operatorname{BrakeDisk.id}$ in $m_1$, triggering a consequential rename in $m_2$. Engineer $B$ renames $\operatorname{Namespace.id}$ in $m_2$, which propagates back to $m_1$. This results in the cycle:
        \[
        a_1 \xrightarrow{\text{consequential rename}} b_1 \qquad \text{and} \qquad b_1 \xrightarrow{\text{consequential rename}} a_1.
        \]
        The presence of this cycle necessitates manual resolution, as no interleaving can satisfy both edges.
    \end{itemize}

    \item \textbf{Topological Sorting as a Serialization Algorithm:}
    \begin{itemize}
        \item Kahn’s algorithm \cite{kahn1962topological} for topologically sorting $G$ is equivalent to finding a \emph{serializable schedule} for the commits. If $G$ is acyclic, the algorithm yields a valid interleaving that:
        \begin{itemize}
            \item Preserves the order of commits within each branch.
            \item Satisfies all inter-branch ordering constraints defined by the footprint overlaps.
        \end{itemize}
        \item This process is described in Section~\ref{sec:formalization:interleaving} but can now be framed formally as a serialization problem. If the algorithm fails to find a valid topological sort due to guard failures (Definition~\ref{def:guard-failure}), the graph is iteratively refined by adding edges (e.g., $\sigma_j \to \sigma_k$) until either a valid sort is achieved or a cycle is detected.
    \end{itemize}
\end{itemize}

\subsection{Correctness Properties}
\label{sec:formalization:properties}

\begin{proposition}[Consistency Preservation]
\label{prop:consistency}
If the merge result $S_M$ is defined by a valid
interleaving, then $\mathrm{Cons}(S_M)$.
\end{proposition}

\begin{proof}[Sketch]
Each replay step computes $\widetilde{S}_k=\pi_S(APP(\widetilde{S}_{k-1},
\sigma_k))$. By Assumption~1, if $\mathrm{Cons}(\widetilde{S}_{k-1})$ and $APP$ is defined, then $\mathrm{Cons}(\widetilde{S}_k)$. The base case $\mathrm{Cons}(S_0)$ holds because the common ancestor results from sequential $APP^*$ application starting from a consistent state. By induction, $\mathrm{Cons}(S_M)$.
\end{proof}

\begin{proposition}[Original-Intent Preservation]
\label{prop:intent}
If the merge returns a conflict-free interleaving~$\sigma$, then no original change from $H_A$ or $H_B$
is silently overwritten. Specifically:
\begin{itemize}
    \item Direct conflicts (original vs.\ original) are detected before replay and require explicit resolution.
    \item The ordering~$\sigma$ is constructed so that, for every consequential overlap (consequential footprint of one commit overlapping the original footprint of another), the original change is replayed \emph{after} the consequential change and overwrites it.
\end{itemize}
\end{proposition}

\begin{proof}[Sketch]
By construction of the dependency graph, every inter-branch edge $a_i \to b_j$ ensures that $a_i$'s consequential change is applied before $b_j$'s original change. Since $\sigma$ is a topological sort of the acyclic graph, all such ordering constraints are satisfied: each original change is the last write to any element-feature pair it targets.
\end{proof}

\paragraph{Summary of guarantees.}
Under Assumption~1 and the scope assumptions in \cref{sec:formalization:scope}, the algorithm has three outcomes: (1)~\emph{success} with a consistent state preserving all original changes (\cref{prop:consistency,prop:intent}); (2)~\emph{direct conflict} (same element-feature, different values, detected before replay); (3)~\emph{consequential conflict} (cycle in dependency graph). Guard failures are resolved by iterative re-ordering; unresolvable failures manifest as cycles.

\subsection{Scope and Limitations}
\label{sec:formalization:scope}

We assume:
\begin{enumerate}[label=(\roman*)]
    \item The consistency preservation function $APP$ correctly restores consistency in the multi-model given a sequence of changes $\tau$ (cf. Assumption 1).
    \item $APP$ does not change between branches, i.e., the same set of Reactions is used on both branches and during merge.
    \item Multi-valued features are treated as sets: independent insertions on both branches are non-conflicting (additive); element ordering is not tracked.
    \item Original changes on each branch are assumed independent of the other branch's state (standard Git-style assumption). The merge does not account for decisions an engineer would have made differently upon seeing concurrent changes.
    \item $CR$ captures inter-model consistency maintained by CPRs; intra-model constraints (multiplicity bounds, OCL invariants) are not part of $CR$. Post-merge metamodel validation is orthogonal.
    \item The guard-failure repair heuristic (adding an edge from the most recent other-branch commit) is sound but not complete: it may fail to find a valid ordering that exists. The implementation falls back to enumerating alternative orderings when the heuristic makes no progress.
\end{enumerate}

\paragraph{The class of consistency relations covered.}
It is worth making explicit which consistency relations the above admits, because the case studies could suggest a narrower class than the definitions require. The consistency relations of the two automotive case studies are value propagations: a target attribute is a function of one or more source attributes (for instance, a safety entry's thermal-load rating computed from a brake disk's diameter and thickness). Nothing in the development depends on this. The definitions in \cref{sec:formalization} treat $CR \subseteq \mathcal{I}$ as an arbitrary predicate over multi-model states and $APP$ as an arbitrary deterministic function restoring it; neither the conflict classification nor the interleaving algorithm inspects the \emph{content} of a change beyond its provenance flag and its footprint, i.e., the set of element-feature pairs it reads or writes. The merge operator is therefore parametric in $CR$: any relation a set of CPRs can restore is admissible, whether or not it is arithmetic.

Case~Study~3 (\cref{sec:evaluation}) instantiates $CR$ with constraints that are \emph{structural} rather than parametric, to show this is not merely a formal possibility: a bijection between architectural components and top-level code packages, which constrains element \emph{existence} and multiplicity rather than attribute values; a correspondence between architectural connectors and authorized edges of the code's package-dependency graph, which constrains a graph's \emph{edge set}; and a naming correspondence in which a component's kind determines, by string transformation on class names, the suffix its package's classes must carry. Because such constraints are expressed over element existence, containment, cross-references, and names, the same features the change metamodel already records, their violations are detected on merge by the same guarded-replay mechanism, with no extension to the operator. What bounds the approach is thus not the arithmetic simplicity of $CR$ but the requirements of Assumption~1 and scope items~(i)--(vi): that $CR$ be restorable by a deterministic $APP$, and that it be an inter-model relation rather than an intra-model one.

%% file: figures/branch-merge-figure.tex
\begin{figure}[t]
\centering

\begin{tikzpicture}[
    vsum/.style={rectangle,draw,fill=black!60,text=white,
    minimum width=1.5cm,minimum height=0.8cm,font=\scriptsize\bfseries},
    merge/.style={rectangle,draw,fill=black!80,text=white,
    minimum width=1.8cm,minimum height=0.9cm,font=\scriptsize\bfseries},
    tau/.style={font=\tiny},
    legendbox/.style={rectangle,draw,rounded corners,font=\scriptsize},
    >=stealth
]

\node[vsum] (A0) at (0,0) {\shortstack{A$_0$\\(base)}};
\node[vsum] (A1) at (2,0) {A$_1$};
\node[vsum] (A2) at (4,0) {A$_2$};
\node[merge] (A3) at (6,0) {\shortstack{Merge\\D}};

\draw[->] (A0) -- node[above,tau] {$\tau_{A_1}$} (A1);
\draw[->] (A1) -- node[above,tau] {$\tau_{A_2}$} (A2);
\draw[->,dashed] (A2) -- node[above,tau] {tbd} (A3);

\node[vsum] (B1) at (2,-1.5) {B$_1$};
\node[vsum] (B2) at (4,-1.5) {B$_2$};

\draw[->] (A0) -- node[left,tau] {$\tau_{B_1}$} (B1);
\draw[->] (B1) -- node[above,tau] {$\tau_{B_2}$} (B2);
\draw[->,dashed] (B2) -- node[right,tau] {tbd} (A3);

\node[legendbox,minimum width=5cm] at (3,-3) {
\begin{tabular}{l}
\textbf{Legend}\\
Each node presents a specific version of VSUM.\\
$\bm{\tau}_{i} =
(\delta_1^{o},\ldots,\delta_n^{o},
\delta_1^{c},\ldots,\delta_m^{c})$\\
Sequence of original ($o$) and consequential ($c$) deltas\\
$\bm{D = merge(base,A_2,B_2)}$
\end{tabular}
};

\end{tikzpicture}

\caption{Branching and merging of VSUM. The merge shall determine $D$.}
\Description{Branching and merging of VSUM states showing main and secondary branches connected by change sequences tau and final merge D.}
\label{fig:branchmerge}
\vspace{-0.5cm}
\end{figure}
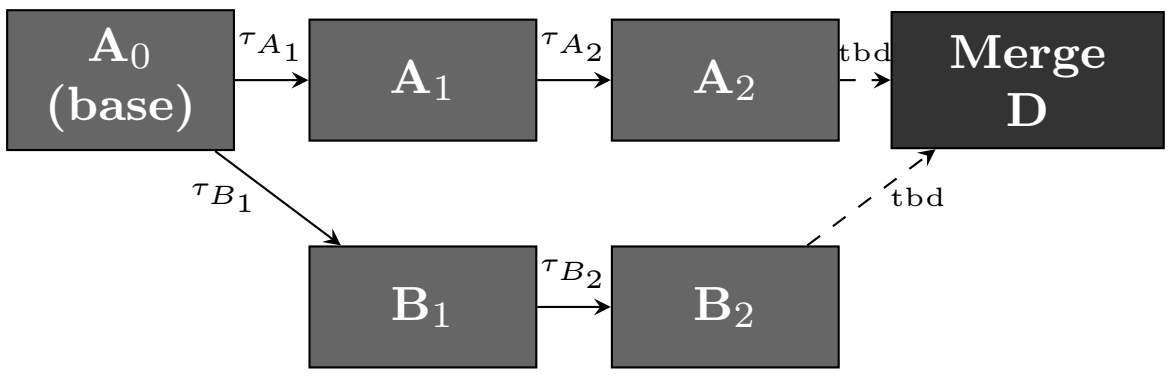

%% file: figures/dependency-graph-figure.tex
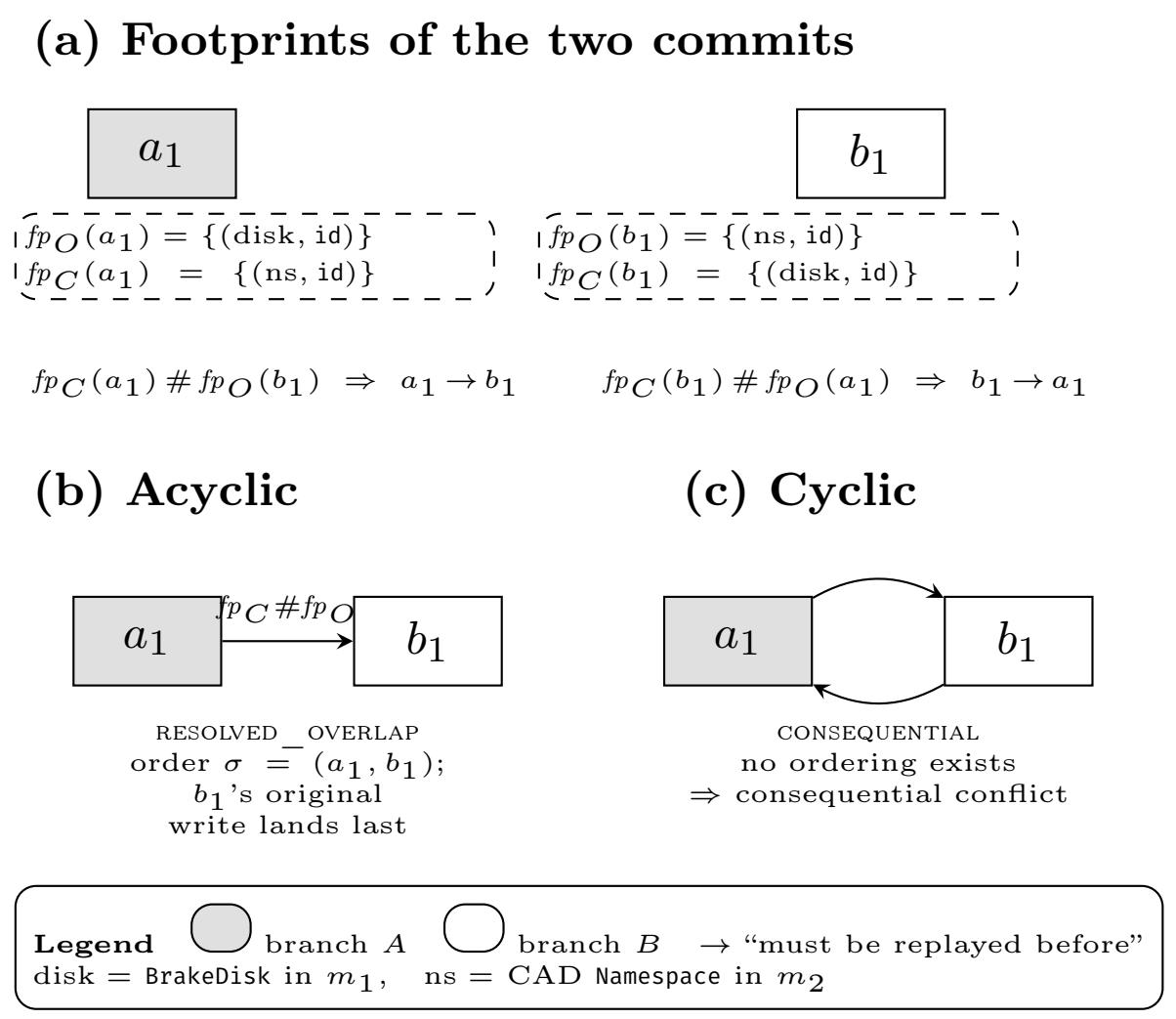
\begin{figure}[t]
\centering

\begin{tikzpicture}[
    commitA/.style={rectangle,draw,fill=black!12,
      minimum width=1.0cm,minimum height=0.6cm,font=\scriptsize\bfseries},
    commitB/.style={rectangle,draw,fill=white,
      minimum width=1.0cm,minimum height=0.6cm,font=\scriptsize\bfseries},
    fpbox/.style={rectangle,draw,dashed,rounded corners,font=\tiny,
      align=left,inner sep=2pt},
    lbl/.style={font=\tiny},
    panel/.style={font=\scriptsize\bfseries},
    >=stealth
]

\node[panel,anchor=west] at (-0.4,2.35) {(a) Footprints of the two commits};

\node[commitA] (ca) at (0.6,1.6) {$a_1$};
\node[commitB] (cb) at (5.4,1.6) {$b_1$};

\node[fpbox,anchor=north west,text width=3.1cm] (fa) at (-0.4,1.2)
  {$\mathit{fp}_O(a_1)=\{(\mathrm{disk},\texttt{id})\}$\\[1pt]
   $\mathit{fp}_C(a_1)=\{(\mathrm{ns},\texttt{id})\}$};
\node[fpbox,anchor=north east,text width=3.1cm] (fb) at (6.4,1.2)
  {$\mathit{fp}_O(b_1)=\{(\mathrm{ns},\texttt{id})\}$\\[1pt]
   $\mathit{fp}_C(b_1)=\{(\mathrm{disk},\texttt{id})\}$};

\node[lbl,anchor=west] at (-0.4,0.05)
  {$\mathit{fp}_C(a_1)\,\#\,\mathit{fp}_O(b_1)\;\Rightarrow\;a_1\!\to\! b_1$
   \qquad
   $\mathit{fp}_C(b_1)\,\#\,\mathit{fp}_O(a_1)\;\Rightarrow\;b_1\!\to\! a_1$};

\node[panel,anchor=west] at (-0.4,-0.7) {(b) Acyclic};

\node[commitA] (aa) at (0.5,-1.7) {$a_1$};
\node[commitB] (ab) at (2.4,-1.7) {$b_1$};
\draw[->] (aa) -- node[above,lbl] {$\mathit{fp}_C\#\mathit{fp}_O$} (ab);

\node[lbl,anchor=north,text width=3.0cm,align=center] at (1.45,-2.15)
  {\textsc{resolved\_overlap}\\ order $\sigma=(a_1,b_1)$;\\ $b_1$'s original write lands last};

\node[panel,anchor=west] at (4.0,-0.7) {(c) Cyclic};

\node[commitA] (ba) at (4.5,-1.7) {$a_1$};
\node[commitB] (bb) at (6.4,-1.7) {$b_1$};
\draw[->] (ba) to[bend left=30] (bb);
\draw[->] (bb) to[bend left=30] (ba);

\node[lbl,anchor=north,text width=3.0cm,align=center] at (5.45,-2.15)
  {\textsc{consequential}\\ no ordering exists\\ $\Rightarrow$ consequential conflict};

\node[rectangle,draw,rounded corners,font=\tiny,align=left,anchor=north west]
  at (-0.4,-3.35)
  {\textbf{Legend}\quad
   \tikz\node[commitA,minimum width=0.4cm,minimum height=0.32cm]{};\ branch $A$\quad
   \tikz\node[commitB,minimum width=0.4cm,minimum height=0.32cm]{};\ branch $B$\quad
   $\rightarrow$ ``must be replayed before''\\
   $\mathrm{disk}=$ \texttt{BrakeDisk} in $m_1$,\quad
   $\mathrm{ns}=$ CAD \texttt{Namespace} in $m_2$};

\end{tikzpicture}

\caption{Construction of the footprint dependency graph. In~(a) each commit
contributes an original and a consequential footprint; an edge is drawn
whenever one commit's \emph{consequential} footprint overlaps the other's
\emph{original} footprint. In~(b) only the $M_1\!\to\!M_2$ direction produces
an overlap, the graph is acyclic, and the topological order places the
original write last so that it is preserved. In~(c) the bidirectional
Reactions produce an overlap in both directions; the cycle means no
interleaving preserves both branches' original intent.}
\Description{Three panels. Panel a shows two commits a1 and b1 with their
original and consequential footprints, and the two overlap conditions that
induce edges. Panel b shows an acyclic graph with a single edge from a1 to b1
and the resulting replay order. Panel c shows a cyclic graph with edges in
both directions between a1 and b1, indicating a consequential conflict.}
\label{fig:depgraph}
\vspace{-0.4cm}
\end{figure}

%% file: sections/05_implementation.tex
\section{Implementation}
\label{sec:implementation}
We implemented the replay-based merge as an extension to the \vitruv framework~\cite{klare2021enabling}. The implementation and evaluation artifacts are available in a replication package~\cite{replicationpackage}. This section describes the main implementation decisions.

\paragraph{Overview.}
\Cref{fig:mergearchitecture} shows where the extension sits. The design rests on one observation: the information the merge needs is available for free \emph{while the engineer works}, and irretrievable afterwards. When the consistency engine propagates an edit, it knows which changes the engineer authored and which it derived itself; once that has been written to disk as a model file, the distinction is gone. The extension therefore does its recording at commit time and its reasoning at merge time.

Accordingly there are two phases. During branch work, nothing about the engineer's workflow changes, edits go through views and the consistency engine runs as normal, but a listener records, per commit, the original changes and the footprint of the consequential ones, and persists them next to the Git commit. During merge, no model states are compared at all: the recorded changes are enough to detect direct conflicts, build the dependency graph of \cref{sec:formalization:interleaving}, and replay each commit back through the same consistency engine, which regenerates the consequential changes rather than merging them.

Only three components are new, changelog capture, conflict detection, and the merge engine itself. Everything else (the consistency engine, views, change recording, UUID assignment, Git) is used unmodified, which is what makes the approach portable in principle to any framework meeting the requirements in \cref{sec:formalization:scope}. \Cref{tab:sourcemap} maps each component to the class that implements it, for readers who wish to follow the description in the source; the prose below does not otherwise assume familiarity with \vitruv{} internals.

\input{figures/merge-architecture-figure}

\begin{table}[t]
\centering
\caption{Where each component of \cref{fig:mergearchitecture} is implemented.}
\label{tab:sourcemap}
\small
\begin{tabular}{ll}
\hline
\textbf{Component} & \textbf{Implementing class} \\
\hline
Changelog capture & \texttt{ChangeLogCapture} \\
Persisted changelog & \texttt{SemanticChangeLog} \\
Direct-conflict detection & \texttt{UuidConflictDetector} \\
Dependency graph & \texttt{CommitDependencyGraph} \\
Replay and interleaving & \texttt{SemanticMergeEngine} \\
Git state extraction & \texttt{GitStateLoader} \\
Git integration & \texttt{GitMergeDriver} \\
Conflict resolution & \texttt{ConflictResolutionProvider} \\
\hline
\end{tabular}
\vspace{-0.3cm}
\end{table}

\subsection{Semantic Changelog Capture}
\label{sec:impl:changelog}

Each commit must record its original changes for later replay, and must record them \emph{with} their provenance. The difficulty is one of timing: provenance is observable only while the consistency engine is running, because that is the only moment at which it is known whether a given change came from the engineer or from a rule. The implementation therefore observes the propagation as it happens, rather than inspecting its result.

A listener attaches to the framework's change-propagation pipeline and intercepts at two points, before and after the consistency rules run:

\begin{itemize}
    \item \textbf{Before rule execution,} the changes present are exactly the ones the engineer authored. These are captured and labelled original.
    \item \textbf{After rule execution,} the additional changes are those the rules derived. Their footprints, the element-feature pairs they wrote, are extracted and stored alongside the original changes. Recording them here is what allows the dependency graph to be built at merge time (\cref{sec:impl:interleaving}) without first replaying anything to find out what the rules would do.
\end{itemize}

Captured changes are also rewritten from the framework's live element references to a self-contained, position-based form, so that a changelog can be deserialized later without the originating session's identity resolver still being available. In \vitruv{} terms, the listener hooks the \texttt{resolveAndApply} $\to$ \texttt{assignIds} $\to$ \texttt{propagateChange} sequence, and the rewriting converts \texttt{EChange<Uuid>} into \texttt{EChange<HierarchicalId>}.

The captured changes are serialized as JSON Data Transfer Objects (DTOs) and stored per commit under \texttt{.vitruvius/semantic-}\-
\texttt{changelogs/<sha>.changelog.json}. Each DTO records the chan\-ge type (e.g., \texttt{ReplaceSingleValuedEAttribute}), the affected element's hierarchical ID and UUID, the EClass name, the feature name, and old/new values. The UUID is the basis for cross-branch identity: elements that existed in the common ancestor carry the same UUID on both branches, while elements created independently receive distinct UUIDs.

\subsection{Element Identity Across Branches}
\label{sec:impl:uuid}

\vitruv assigns each model element a UUID persisted in a Git-tracked file (\texttt{vsum/uuid.uuid}). Elements present at the branch point carry identical UUIDs on both branches; elements created independently receive different UUIDs and never conflict. This stability is the foundation for cross-branch conflict detection.

\subsection{Merge Engine}
\label{sec:impl:interleaving}

The \texttt{SemanticMergeEngine} implements the interleaving algorithm from \cref{sec:formalization:interleaving}. Given branches $A$ and $B$ with $p=|H_A|$ and $q=|H_B|$ commits respectively, the merge proceeds as follows.

\paragraph{State extraction and direct conflict detection.}
The engine uses JGit's \texttt{TreeWalk} to extract model files, UUID mappings, and changelogs from the base, $A$-head, and $B$-head commits into temporary directories. Changelog DTOs from both branches are then compared by UUID and feature name to detect direct conflicts (same element-feature modified by both branches with different values). This detection operates on DTOs without loading model instances, making it efficient.

\paragraph{Footprint computation.}
Direct footprints are extracted from the changelog DTOs (UUID and feature name per change). Consequential footprints are captured at commit time by the \texttt{ChangeLogCaptu\-re} listener (\cref{sec:impl:changelog}) alongside the original changes and stored in the changelog JSON.

\paragraph{Dependency graph construction and iterative refinement.}
A dependency graph is constructed with intra-branch ordering edges and inter-branch edges from footprint overlaps (\cref{sec:formalization:interleaving}). A cycle yields a consequential conflict; otherwise, Kahn's algorithm produces a topological ordering executed on a copy of the base state. Each commit's changes are deserialized into live \texttt{EChange} objects and applied via a \texttt{ChangeRecordingView}; \texttt{commitChanges()} propagates through \vitruv's Reaction engine, producing consequential changes as during normal operation. Elements are resolved by HierarchicalId with UUID-based fallback.

Actual consequential footprints captured during execution are compared to estimates; new footprints trigger iterative refinement (\cref{sec:formalization:interleaving}).

\paragraph{Conflict resolution.}
When conflicts are detected (direct or consequential), a \texttt{ConflictResolutionProvider} is invoked. The implementation supports programmatic resolution (\texttt{chooseAllOurs()}, \texttt{chooseAllTheirs()}) and per-conflict custom resolution.

\subsection{Integration with Git}

The merge engine operates on a standard Git repository. Branch management (creation, switching, deletion) is handled via JGit. Git hooks (\texttt{post-checkout}, \texttt{pre-commit}, \texttt{post-commit}) trigger \vsum reload after branch switches, validate consistency before commits, and persist semantic changelogs after commits.

To integrate with Git's native merge workflow, we implement a custom merge driver~\cite{gitattributes}.
Since our merge requires the full \vsum rather than individual files, the driver runs the complete semantic merge on first invocation and caches the resulting model state. Subsequent per-file invocations copy the corresponding merged file from the cache. If the merge succeeds, Git creates the merge commit automatically; if blocking conflicts are detected, the driver signals failure and the merge is aborted.

%% file: figures/merge-architecture-figure.tex
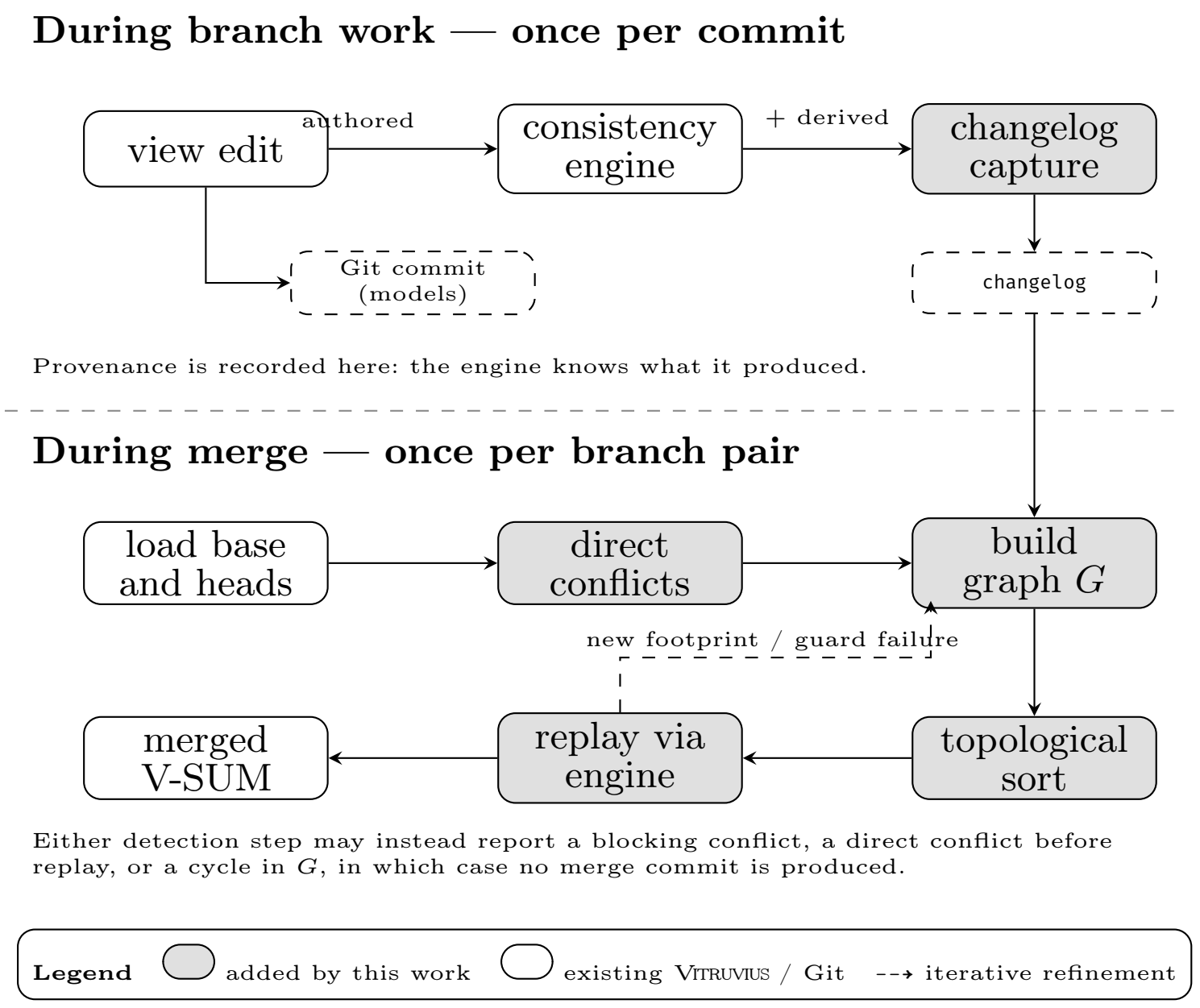
\begin{figure}[t]
\centering

\begin{tikzpicture}[
    process/.style={rectangle,draw,rounded corners,fill=white,
      text width=1.86cm,minimum height=0.62cm,inner sep=2pt,
      font=\scriptsize,align=center},
    ours/.style={process,fill=black!12},
    store/.style={rectangle,draw,dashed,rounded corners,fill=white,
      text width=1.86cm,minimum height=0.5cm,inner sep=2pt,
      font=\tiny,align=center},
    lane/.style={font=\scriptsize\bfseries},
    note/.style={font=\tiny,align=center},
    lnote/.style={font=\tiny,align=left},
    >=stealth
]

\node[lane,anchor=west] at (-0.5,3.20) {During branch work --- once per commit};

\node[process]  (edit)  at (1.05,2.25) {view edit};
\node[process]  (prop)  at (4.45,2.25) {consistency engine};
\node[ours]  (cap)   at (7.85,2.25) {changelog capture};

\draw[->] (edit) -- (prop);
\draw[->] (prop) -- (cap);
\node[note,anchor=south] at (2.30,2.30) {authored};
\node[note,anchor=south] at (6.15,2.30) {$+$ derived};

\node[store] (git)  at (2.75,1.15) {Git commit (models)};
\node[store] (json) at (7.85,1.15) {\texttt{changelog}};

\draw[->] (cap) -- (json);
\draw[->] (edit.south) |- (git.west);

\node[lnote,anchor=west] at (-0.5,0.45)
  {Provenance is recorded here: the engine knows what it produced.};

\draw[dashed,black!45] (-0.6,0.10) -- (9.0,0.10);

\node[lane,anchor=west] at (-0.5,-0.25) {During merge --- once per branch pair};

\node[process] (load)  at (1.05,-1.15) {load base and heads};
\node[ours] (direct)at (4.45,-1.15) {direct conflicts};
\node[ours] (graph) at (7.85,-1.15) {build graph $G$};

\draw[->] (load) -- (direct);
\draw[->] (direct) -- (graph);
\draw[->] (json) -- (graph);

\node[process] (out)   at (1.05,-2.75) {merged \vsum};
\node[ours] (replay)at (4.45,-2.75) {replay via engine};
\node[ours] (sort)  at (7.85,-2.75) {topological sort};

\draw[->] (graph) -- (sort);
\draw[->] (sort) -- (replay);
\draw[->] (replay) -- (out);

\draw[->,dashed] (replay.north) -- ++(0,0.45) -| (7.0,-1.46);
\node[note,anchor=south] at (5.70,-2.0) {new footprint / guard failure};

\node[lnote,anchor=west] at (-0.5,-3.55)
  {Either detection step may instead report a blocking conflict, a direct conflict before\\
   replay, or a cycle in $G$, in which case no merge commit is produced.};

\node[rectangle,draw,rounded corners,font=\tiny,align=left,anchor=north west]
  at (-0.5,-4.15)
  {\textbf{Legend}\quad
   \tikz\node[ours,text width=0.35cm,minimum height=0.28cm,inner sep=1pt]{};\ added by this work\quad
   \tikz\node[process,text width=0.35cm,minimum height=0.28cm,inner sep=1pt]{};\ existing \vitruv{} / Git\quad
   $\dashrightarrow$ iterative refinement};

\end{tikzpicture}

\caption{How the merge interacts with \vitruv{} and Git. During branch work the
consistency engine runs as usual; the added changelog capture records, per
commit, which changes the engineer authored and which the engine derived, and
persists them next to the Git commit. During merge, nothing is compared
state-wise: the recorded changelogs drive direct-conflict detection, the
construction of the dependency graph~$G$, and the replay of each commit back
through the same consistency engine, which yields the merged \vsum{} and its
merge commit. Only the shaded steps are new.}
\Description{Two lanes. The upper lane shows, during branch work, a view edit
flowing into the consistency engine and then into changelog capture, which
writes a per-commit changelog alongside the Git commit. The lower lane shows,
during merge, loading the base and both branch heads, direct conflict
detection, building the dependency graph, topological sorting, replaying each
commit through the consistency engine, and producing the merged V-SUM, with a
dashed refinement loop from replay back to graph construction.}
\label{fig:mergearchitecture}
\vspace{-0.4cm}
\end{figure}

%% file: sections/06_evaluation.tex
\section{Evaluation}
\label{sec:evaluation}

We evaluate our approach on three case studies from different engineering domains and compare against EMF~Compare~\cite{emfcompare} as a state-based baseline. The evaluation addresses three questions:
\begin{description}
    \item[RQ1] Does provenance-aware conflict classification reduce false-positive blocking conflicts compared to state-based merge?
    \item[RQ2] Does the conflict reduction advantage persist across longer branch histories, varying overlap densities, and different Reaction densities?
    \item[RQ3] How does merge performance scale with model size and branch history length?
\end{description}

\subsection{Experimental Setup}

\paragraph{Case study domains.}
We evaluate on three case studies, each using a three-model \vsum maintained by \vitruv. The first two are from automotive engineering and maintain consistency relations that are, in essence, value propagation over numeric or nominal attributes; the third is from software engineering and maintains \emph{structural} consistency relations, so that the evaluation is not confined to one domain or to one kind of consistency relation.

\emph{Case Study~1: Automotive Brake System.}
The brake system case study derived from Hagel et al.'s work~\cite{hagel2025brakecasestudy} uses three metamodels ($M_1$: Brake\-system, $M_2$: CAD, $M_3$: Safety) linked by three sets of \vitruv Reactions ($M_1 \leftrightarrow M_2$, $M_1 \to M_3$), as described in \cref{sec:example}.

\emph{Case Study~2: Automated Driving System (MobSTr).}
Based on the MobSTr dataset~\cite{mobstr_2021_06,mobstr_example} from the WATERS FMTV~2019 industrial challenge, this case study models a safety-critical lane-following system with three metamodels: $M_1$:~Architecture (components with typed ports and connections), $M_2$:~Amalthea (tasks, runnables, labels for timing analysis), and $M_3$:~FMEA (component failure models with classified failure modes). The models are linked by three Reaction sets: $M_1 \leftrightarrow M_2$ (bidirectional) and $M_1 \to M_3$ (unidirectional).

\emph{Case Study~3: Code--Architecture Consistency (TEAMMATES).}
TEAMMATES~\cite{goh2011teammates,teammates_repo} is a mature open-source web application for managing peer feedback in student team projects, maintained for more than a decade. We selected it because its architecture is documented as a component-and-connector model (\texttt{highlevelArchitecture.puml}) whose layering rules are also encoded as \emph{executable} constraints: the project ships an ArchUnit test suite of 40~rules that mechanically relate the Java package structure to the documented architecture (for example, \texttt{ui} must not depend on \texttt{storage}, and every \texttt{*Logic} class may access only its correspondingly named \texttt{*Db} class). The consistency relation we maintain is therefore not one we devised to fit the approach, but one the project already curates. We model the system with three metamodels: $M_1$:~Architecture (components with a \emph{kind}, and the connectors that authorize dependencies between them), $M_2$:~Code (top-level Java packages, the classes they declare, and declared package-to-package dependencies), and $M_3$:~Architecture rules (the derived dependency and naming-convention rules, i.e., the model-level counterpart of \texttt{ArchitectureTest.java}), linked by $M_1 \leftrightarrow M_2$ (bidirectional) and $M_1 \to M_3$ (unidirectional) Reaction sets.

Unlike the two automotive case studies, the consistency relations here are \emph{structural} rather than value-propagating:
(CR1)~each component corresponds to exactly one top-level package of the same name (existence and multiplicity);
(CR2)~a connector $A\!\rightarrow\!B$ corresponds to the permission for a package dependency $\mathit{pkg}_A\!\rightarrow\!\mathit{pkg}_B$ (link topology);
(CR3)~a component's kind fixes the class-name suffix its package must use (\texttt{STORAGE}$\rightarrow$\texttt{*Db}, \texttt{LOGIC}$\rightarrow$\texttt{*Logic}), a naming correspondence realized by string transformation that mirrors the project's own \texttt{Logic}$\rightarrow$\texttt{Db} rule; and
(CR4)~every component and connector derives its architecture rule in $M_3$, and deleting a component cascades to deleting its derived rules.
CR4 captures the discipline whose absence let the real TEAMMATES rules go stale after the project's migration from a NoSQL (Datastore/Objectify) to a SQL (Hibernate/PostgreSQL) backend: two rules still reference the deleted \texttt{storage.search} package and pass vacuously. $M_2$ is an abstracted code model, packages, classes, declared dependencies, and name suffixes, not a full Java abstract syntax tree, which matches the abstraction level at which the project's own architecture constraints are expressed.

Each scenario creates a Git repository with a base commit, then introduces changes on two branches, and merges them.

\paragraph{Scenarios.}
For each case study, we selected representative scenarios covering both conflict types (direct and consequential; \cref{sec:formalization:conflicts}) and the interleaving merge: five scenarios for Case~Study~1 (\textsc{clean}, \textsc{resolved\_overlap}, \textsc{direct}, \textsc{consequential}, \textsc{multi\_com\-mit}), four for Case~Study~2 (\textsc{clean}, \textsc{resolved\_overlap}, \textsc{direct}, \textsc{consequential}), and five for Case~Study~3 (\textsc{clean}, \textsc{resolved\_overlap}, \textsc{direct}, \textsc{consequential}, \textsc{delete\_modify}). The Case~Study~3 scenarios are edit pairs taken from the project's real NoSQL-to-SQL migration rather than invented for the evaluation. Each scenario is implemented as an automated JUnit test. For RQ2, we additionally generated 130~scenarios automatically using a parameterized scenario generator with controlled randomness (five deterministic seeds per configuration) on Case~Study~1.

\paragraph{Baseline.}
EMF~Compare performs state-based three-way comparison on XMI model files. For each scenario, we extract the base, source, and target model states from Git, run EMF~Compare's three-way comparison independently on each model file ($M_1$, $M_2$, $M_3$), and count the reported conflicts. EMF~Compare operates without knowledge of Reactions or the distinction between original and consequential changes.

EMF~Compare itself classifies each conflict it reports as either \textsc{real} or \textsc{pseudo}: a \textsc{pseudo} conflict is one in which both branches made the \emph{same} change to the same location, so it can be resolved automatically and requires no human decision. We count only \textsc{real} conflicts, and treat a scenario as a successful EMF~Compare merge exactly when no \textsc{real} conflict remains. The baseline figures we report are therefore a \emph{lower bound} on the number of locations EMF~Compare would surface: counting \textsc{pseudo} conflicts as well would increase the baseline counts and widen the reported reduction. We make this explicit because the comparison is only meaningful if both sides are counted at the granularity of decisions a developer actually has to make.

\paragraph{Baseline selection.}
We chose EMF~Compare as baseline because it is the de facto standard for EMF model merging~\cite{brun2008model,sharbaf2023conflict}. EML~\cite{kolovos2006merging} was considered but excluded as it is a two-way merge language without native three-way support; adapting it would produce a contrived setup not representative of how EML is used in practice.

We also considered the operation-based tool EMFStore~\cite{koegel2010emfstore} as a delta-based baseline but excluded it: although EMFStore records edit operations, its merge model does not distinguish original from consequential changes. The observed conflict reduction stems from this provenance distinction, not from using recorded operations per se. EMFStore would differ from EMF~Compare in conflict granularity but not with respect to the provenance gap our approach addresses.

\subsection{RQ1 Results: Conflict Classification}

\Cref{tab:comparison} summarizes the results for all three case studies.

\begin{table}[t]
\centering
\caption{Merge results: Number of conflicts (C) reported by \vitruv (ours) vs.\ EMF~Compare for all three case studies and time of ours in ms}
\label{tab:comparison}
\small

\begin{tabular}{llccr}
\hline
\textbf{Label} & \textbf{Category} & \textbf{Ours} & \textbf{EMF~C} & \textbf{Time} \\
\hline
\multicolumn{5}{l}{\textit{Case Study~1: Brake System}} \\
\textsc{clean}             & Clean merge               & 0 & 0 & $16.0 \pm 2.7$ \\
\textsc{resolved\_overlap} & Conseq.\ overlap resolved & 0 & 1 & $10.7 \pm 1.5$ \\
\textsc{direct}            & Direct conflict           & 1 & 3 & $9.4 \pm 0.3$ \\
\textsc{consequential}     & Conseq.\ conflict (cycle) & 2 & 5 & $4.8 \pm 1.2$ \\
\textsc{multi\_commit}     & Interleaving              & 0 & 2 & $11.6 \pm 1.7$ \\
\cmidrule{2-5}
                            & \textbf{Subtotal}        & \textbf{3} & \textbf{11} & \\
\hline
\multicolumn{5}{l}{\textit{Case Study~2: Automated Driving (MobSTr)}} \\
\textsc{clean}             & Clean merge               & 0 & 0 & $18.2 \pm 2.3$ \\
\textsc{resolved\_overlap} & Conseq.\ overlap resolved & 0 & 1 & $9.9 \pm 1.0$ \\
\textsc{direct}            & Direct conflict           & 1 & 6 & $8.9 \pm 1.1$ \\
\textsc{consequential}     & Conseq.\ conflict (cycle) & 2 & 6 & $3.5 \pm 0.2$ \\
\cmidrule{2-5}
                            & \textbf{Subtotal}        & \textbf{3} & \textbf{13} & \\
\hline
\multicolumn{5}{l}{\textit{Case Study~3: Code--Architecture (TEAMMATES)}} \\
\textsc{clean}             & Clean merge               & 0 & 0 & $14.5 \pm 1.2$ \\
\textsc{resolved\_overlap} & Conseq.\ overlap resolved & 0 & 1 & $9.8 \pm 0.5$ \\
\textsc{direct}            & Direct conflict           & 1 & 4 & $9.2 \pm 0.5$ \\
\textsc{consequential}     & Conseq.\ conflict (cycle) & 2 & 4 & $3.7 \pm 0.3$ \\
\textsc{delete\_modify}    & Delete/modify             & 1 & 4 & $13.2 \pm 1.1$ \\
\cmidrule{2-5}
                            & \textbf{Subtotal}        & \textbf{4} & \textbf{13} & \\
\hline
                            & \textbf{Total}           & \textbf{10} & \textbf{37} & \\
\hline
\end{tabular}

\smallskip
\raggedright\footnotesize
C = conflicts (counted per element-feature pair). Time = wall-clock merge time (mean $\pm$ stdev, 5 runs after one discarded warmup), measured for all three case studies on the same machine (\cref{sec:evaluation}).
\vspace{-0.5cm}
\end{table}

\paragraph{Case Study~1: Brake System.}

\emph{\textsc{clean}.}
Engineer~A changes \texttt{BrakeDi\-sk .diameterInMM} in~$m_1$; the $M_1 \to M_2$ and $M_1 \to M_3$ Reactions derive updates to the CAD parameter and safety thermal load rating, respectively. Engineer~B adds a new \texttt{ABSSensor} component. The changes are on different elements, so both approaches report zero conflicts.

\emph{\textsc{resolved\_overlap}.}
Engineer~A changes the brake disk diameter in~$m_1$; the $M_1 \to M_2$ Reaction derives a CAD Diameter parameter update. Engineer~B directly sets that same CAD Diameter to a different value in~$m_2$. EMF~Compare sees two different values for the same parameter and reports one blocking conflict. Our approach recognizes that A's CAD change is consequential (not original). The dependency graph is acyclic: A's consequential footprint overlaps B's original footprint, which orders $a_1$ before $b_1$, so B's original write is the last to land on that parameter and its intent is preserved.

\emph{\textsc{direct}.}
Both engineers change \texttt{BrakeDisk.diameterInMM} to different values. Our approach detects one blocking conflict on the root cause in~$m_1$. EMF~Compare reports three conflicts: one in~$m_1$, one in~$m_2$ (consequential CAD parameter), and one in~$m_3$ (consequential safety thermal load rating). The two additional EMF~Compare conflicts are on consequential model elements that would be regenerated correctly once the root conflict is resolved.

\emph{\textsc{consequential}.}
Engineer~A renames \texttt{BrakeDisk.id} in~$m_1$, which triggers Reaction cascades updating the CAD namespace ID in~$m_2$ and the safety entry ID in~$m_3$. Engineer~B renames the CAD namespace ID in~$m_2$, which triggers the $M_2 \to M_1$ Reaction deriving a component ID change in~$m_1$. The bidirectional Reactions create a cycle in the dependency graph: A's consequential footprint (namespace ID in~$m_2$) overlaps B's original footprint, and B's consequential footprint (component ID in~$m_1$) overlaps A's original footprint. No interleaving can preserve both branches' original intent. Our approach identifies two blocking consequential conflicts (one per overlapping element-feature pair in the cycle); EMF~Compare reports five conflicts across all three models.

\emph{\textsc{multi\_commit}.}
Branch~A has two commits: $a_1$ changes the brake disk diameter (the $M_1 \to M_3$ Reaction derives \texttt{thermalLoadRating}), and $a_2$ directly sets \texttt{frictionArea} in~$m_3$. Branch~B has two commits: $b_1$ directly sets \texttt{thermalLoadRating} in~$m_3$, and $b_2$ changes the brake pad height (the $M_1 \to M_3$ Reaction derives \texttt{frictionArea}).
The dependency graph has edges $a_1 \to b_1$ and $b_2 \to a_2$ but no cycle, so topological sort yields the interleaving $[a_1, b_1, b_2, a_2]$, which avoids both overlaps. Our approach merges successfully; EMF~Compare reports two conflicts.

\paragraph{Case Study~2: Automated Driving System (MobSTr).}

The MobSTr case study models a lane-following system where two engineers work on the architecture concurrently. The base state contains components GPS (with an output port) and Localization (with two ports), linked by a data-flow connection that triggers shared-label management in the Amalthea timing model.

\emph{\textsc{clean}.}
Engineer~A renames \texttt{GPS} to \texttt{GNSS} in~$m_1$, triggering consequential updates across all three models. Engineer~B adds a \texttt{Camera} component. Disjoint changes; both approaches report zero conflicts.

\emph{\textsc{resolved\_overlap}.}
Engineer~A renames \texttt{GPS} to \texttt{GNSS} in~$m_1$, cascading to the FMEA \texttt{componentName} in~$m_3$. Engineer~B directly sets that field to \texttt{GPS-v2}. The unidirectional $M_1 \to M_3$ link yields an acyclic dependency graph; interleaving replays A first, so B's original edit lands last and is preserved. EMF~Compare reports one conflict.

\emph{\textsc{direct}.}
Both engineers rename \texttt{GPS} to different values. Our approach detects one blocking conflict on the root cause in~$m_1$; EMF~Compare reports six across all three models due to Reaction cascades.

\emph{\textsc{consequential}.}
Engineer~A renames \texttt{GPS} in~$m_1$ (cascading to Task name in~$m_2$); Engineer~B renames the same Task in~$m_2$ (cascading to component name in~$m_1$). The bidirectional Reactions create a cycle analogous to Brake \textsc{consequential}. Our approach reports two blocking consequential conflicts; EMF~Compare reports six.

\paragraph{Case Study~3: Code--Architecture Consistency (TEAMMATES).}

The base state contains the four production components \texttt{ui}, \texttt{logic}, \texttt{storage} and \texttt{search}, together with the connector that authorizes the Logic-to-Storage dependency. Each scenario pairs two edits drawn from the NoSQL-to-SQL migration.

\emph{\textsc{clean}.}
Engineer~A renames the \texttt{storage} component in~$m_1$, which cascades to its package in~$m_2$ and its derived rule in~$m_3$. Engineer~B adds the \texttt{liquibase} component the migration introduced. The changes are on disjoint elements, so both approaches report zero conflicts.

\emph{\textsc{resolved\_overlap}.}
Engineer~A renames the \texttt{storage} component in~$m_1$; the $M_1 \to M_3$ Reaction derives an update to the corresponding naming rule in~$m_3$. Engineer~B hand-edits that same rule, the way the project's real rules drifted. Because $M_1 \to M_3$ is unidirectional, the dependency graph is acyclic and interleaving replays A first, so B's original edit lands last and is preserved. EMF~Compare reports one conflict.

\emph{\textsc{direct}.}
Both engineers rename the \texttt{storage} component to different names, the genuine naming question raised when \texttt{InstructorsDb} and \texttt{StudentsDb} were merged into \texttt{UsersDb}. Our approach detects one blocking conflict on the root cause in~$m_1$; EMF~Compare reports four across the three models, the three additional ones on consequential elements that would be regenerated once the root conflict is resolved.

\emph{\textsc{consequential}.}
Engineer~A renames the \texttt{storage} component in~$m_1$, whose Reaction derives the corresponding package rename in~$m_2$; Engineer~B renames that same package directly in~$m_2$, whose Reaction derives a component rename in~$m_1$. As in the two automotive case studies, the bidirectional Reactions create a cycle in the footprint dependency graph. Our approach reports two blocking consequential conflicts; EMF~Compare reports four.

\emph{\textsc{delete\_modify}.}
Engineer~A removes the \texttt{search} component, as the migration did, cascading to its package in~$m_2$ and its derived rule in~$m_3$; Engineer~B renames the same component. Both changes are original, so the deletion and the modification target the same element and our approach reports one blocking direct conflict. EMF~Compare reports four, the three additional ones on the cascade-deleted consequential elements.

\paragraph{Summary (RQ1).}
Across all three case studies (14~scenarios total), our approach reports \textbf{10~blocking conflicts} (counted per element-feature pair): one \textsc{direct} conflict and two \textsc{consequential} conflicts per case study (one per overlapping element-feature pair in the dependency graph cycle), plus the one \textsc{delete\_modify} conflict in Case~Study~3. EMF~Compare reports \textbf{37~blocking conflicts} in total. The pattern is consistent across domains: \vitruv reduces false-positive blocking conflicts by 8~in the brake system, by 10~in the driving system, and by 9~in TEAMMATES. Case~Study~3 shows the reduction is not tied to value-propagation consistency relations: the same reduction arises where the relations constrain element existence, dependency topology, and naming. In every case study the genuine \textsc{direct} and cyclic conflicts are retained, so the reduction does not come at the cost of under-reporting real conflicts.

Beyond conflict counts, each scenario is verified against a per-scenario oracle that asserts: (i)~the expected merge outcome, (ii)~preservation of original changes from both branches, (iii)~correct regeneration of consequential changes by Reactions, and (iv)~cross-model consistency via custom validators that check the consistency relation $CR$ directly (e.g., component IDs in~$m_1$ match namespace IDs in~$m_2$ and safety entry IDs in~$m_3$; \texttt{thermalLoadRating} equals the formula applied to the merged diameter and thickness). For Case~Study~3 the validator checks the \emph{structural} invariants CR1--CR4 instead: that every component still corresponds to exactly one same-named package and vice versa, that every declared package dependency is authorized by a connector, that class-name suffixes match their component's kind, and that no architecture rule survives the deletion of the component it was derived from. These validators are independent of the merge mechanism and verify that the merged state satisfies inter-model invariants. All successful merges satisfy their oracle.

\subsection{Coverage of the Conflict Space}
\label{sec:evaluation:coverage}

The scenarios above were selected to be representative, which invites the question of what they are representative of. This subsection makes the coverage argument explicit by classifying the space of possible cross-branch interactions and locating each scenario in it.

Two dimensions determine how the algorithm treats an interaction, and they are exactly the two the formalization is built on (\cref{sec:formalization:conflicts}): the \emph{provenance pair} of the two overlapping changes, and whether the resulting dependency graph is \emph{cyclic}. \Cref{tab:coverage} enumerates the resulting cells.

\begin{table}[t]
\centering
\caption{The conflict space by provenance pair and graph shape, with the scenarios that instantiate each reachable cell.}
\label{tab:coverage}
\small
\begin{tabular}{llll}
\hline
\textbf{Provenance pair} & \textbf{Graph} & \textbf{Outcome} & \textbf{Instantiated by} \\
\hline
original--original     & ---      & direct conflict & \textsc{direct} (CS1--3), \textsc{delete\_modify} (CS3) \\
consequential--original & acyclic & auto-resolved   & \textsc{resolved\_overlap} (CS1--3), \textsc{multi\_commit} (CS1) \\
consequential--original & cyclic  & conseq.\ conflict & \textsc{consequential} (CS1--3) \\
consequential--consequential & --- & no constraint  & F6 (25 generated scenarios) \\
disjoint               & acyclic  & clean merge     & \textsc{clean} (CS1--3) \\
\hline
\end{tabular}
\vspace{-0.3cm}
\end{table}

Three of the cells that a naive enumeration would produce are unreachable, and the reasons are instructive rather than incidental.

\emph{Original--original conflicts are insensitive to graph shape.} They are detected statically from the recorded changes before any graph is built (\cref{sec:formalization:conflicts}), so the cyclic and acyclic sub-cases collapse: the conflict is reported either way. This is why the row carries no graph column.

\emph{Consequential--consequential overlaps induce no edge at all.} An edge requires one commit's \emph{consequential} footprint to meet the other's \emph{original} footprint. Where both sides are consequential, neither change is being preserved, both are regenerated by replay, so there is nothing to order and no conflict to report. This is not a gap in the classification but the mechanism by which the reduction is obtained, and Family~6 of RQ2 instantiates it directly: across 25 generated scenarios in which both branches modify complementary Reaction-triggering attributes of the same elements, every merge succeeds with zero conflicts while EMF~Compare reports between two and eight.

\emph{The provenance pair is symmetric.} A ``consequential--original'' and an ``original--consequential'' overlap are the same relation read from the two branches' perspectives; what distinguishes the cases is not which branch is which but \emph{how many} of the two directions are present. One direction yields an ordering constraint the topological sort discharges; both yield the cycle that constitutes a consequential conflict. The two rows in \cref{tab:coverage} are therefore one-direction and two-direction, not $A$-first and $B$-first.

Orthogonally, the atomic operations of \cref{tab:operations} are exercised as follows: $\mathsf{set}$ by every value-propagation scenario, $\mathsf{new}$ by \textsc{clean} (adding a component on one branch while the other edits), and $\mathsf{del}$, with the containment cascade of \cref{sec:formalization:footprints}, by \textsc{delete\_modify}. Insertion and removal on multi-valued features ($\mathsf{ins}$, $\mathsf{rem}$) occur within these scenarios wherever an element is added to or removed from a containment feature, but we do not isolate them as separate scenarios.

\paragraph{What this does not cover.}
Two limitations follow from the table rather than from the individual scenarios. First, a scenario exhibiting an original--original conflict \emph{and} a cycle simultaneously is not among our cases; the algorithm would report the direct conflict first, so we expect no new behaviour, but we have not measured it. Second, the classification is over pairs of overlapping changes, whereas a merge may contain several overlaps at once; \textsc{multi\_commit} and the 130~generated scenarios of RQ2 exercise such combinations, but not systematically by cell. A fully crossed design over operations, provenance pairs and graph shapes remains future work; the argument here is that the cells the algorithm distinguishes are covered, not that every combination of them is.

\subsection{RQ2 Results: Robustness of Reduction}

RQ1 demonstrates conflict reduction on hand-crafted scenarios. To assess whether this advantage holds under more varied and realistic conditions, we automatically generate 130~merge scenarios across six experimental families, each varying one factor while controlling the others. All scenarios use five seeds for variance estimation.

\paragraph{Scenario generator.}
A parameterized generator creates Git repositories with controlled base states (2--10 brake components), then produces multi-commit branch histories (1--25~commits per branch) with controlled element-level overlap between branches (0--100\%) and controlled fraction of changes targeting Reaction-triggering attributes (20--80\%). Each commit contains 1--3~randomly selected model operations (20\% additive, 80\% modifications of existing elements from the branch's focus set). Overlap is instantiated by selecting overlapping subsets of base elements as each branch's focus set. Both our merge and EMF~Compare are run on each generated scenario. Conflict counts are averaged over five seeds; we report mean and standard deviation.

The six families are: F1~varies history length ($k \in \{1,3,5,10,25\}$), F2~varies overlap density ($0.0$--$1.0$), F3~varies Reaction density ($0.2$--$0.8$), F4~varies base state size ($2$--$10$ components), and F5~varies overlap$\,{\times}\,$Reaction density with interleaving enabled. F6~uses a different generation strategy: both branches modify \emph{complementary} Reaction-triggering attributes of the same elements (e.g., diameter vs.\ thickness of the same BrakeDisk), creating overlapping consequential footprints without direct conflicts. Full generator details are documented in the replication package~\cite{replicationpackage}.

\paragraph{Family~1: History length scaling.}
\Cref{tab:trackb-history} shows how conflict counts change as branch histories grow from 1 to 25~commits per branch, with fixed base size ($n=5$), overlap ($0.3$), and Reaction density ($0.5$).

\begin{table}[t]
\centering
\caption{Conflict reduction vs.\ branch history length ($n\!=\!5$, overlap$\,=\!0.3$, reaction$\,=\!0.5$).}
\label{tab:trackb-history}
\small
\begin{tabular}{rccc}
\hline
\textbf{$k$} & \textbf{Ours: C} & \textbf{EMF~C: C} & \textbf{Reduction} \\
\hline
1  & 0.4 & 0.8  & 0.4 \\
3  & 0.4 & 2.6  & 2.2 \\
5  & 3.8 & 6.6  & 2.8 \\
10 & 4.0 & 10.6 & 6.6 \\
25 & 7.2 & 15.2 & 8.0 \\
\hline
\end{tabular}

\smallskip
\raggedright\footnotesize $k$ = commits per branch. C = conflicts. Values are averages over 5~seeds. Reduction = EMF~C $-$ Ours~C.
\vspace{-0.5cm}
\end{table}

As branch histories grow, EMF~Compare's conflict count increases rapidly (from 0.8 to 15.2) because each additional commit generates consequential changes that accumulate as false-positive conflicts. Our approach's conflict count grows more slowly (0.4 to 7.2), and the \emph{reduction gap widens}: from 0.4 at $k\!=\!1$ to 8.0 at $k\!=\!25$.

\paragraph{Families~2--6.}
F2 (overlap density): increasing element-level overlap (0.0--1.0) increases conflicts for both approaches, but EMF~Compare grows faster (up to 6.6 vs.\ our 4.2 at overlap$\,{=}\,0.6$).
F3 (Reaction density): increasing the fraction of Reaction-triggering changes (0.2--0.8) amplifies the advantage (average reduction 3.5), confirming that more Reactions produce more false-positive consequential conflicts.
F4 (base state size): varying components from 2 to 10 yields an average reduction of 3.0.
F5 (interleaving): across 30~scenarios with varying overlap and Reaction density, our approach reports 3.3~conflicts on average vs.\ EMF~Compare's 7.1.
F6 (consequential overlap): across 25~scenarios where both branches modify complementary attributes of the same elements, our approach merges all successfully with zero conflicts while EMF~Compare reports 2--8 per scenario.
Detailed per-family tables are in the replication package~\cite{replicationpackage}.

\paragraph{Summary (RQ2).}
Across 130~automatically generated scenarios, our approach reports on average 2.6~conflicts compared to EMF~Compare's 6.6 (an average reduction of 4.0). This advantage persists and \emph{increases} with longer branch histories and greater element overlap, and remains stable across varying Reaction densities and base state sizes. Family~6 demonstrates that when consequential overlaps are the only source of cross-branch interference, our approach achieves a 100\% success rate with zero blocking conflicts. All 27~successful merges across all families (2 from Families~1--5, 25 from Family~6) pass $CR$ validation, confirming that the merged states satisfy inter-model consistency.

\subsection{RQ3 Results: Performance and Scalability}

To assess whether replay-based merge remains practical as models and branch histories grow, we measure wall-clock merge time under controlled scaling of two factors: model size (number of components in the multi-model) and branch history length (number of replayed commits).

\paragraph{Setup.}
A parameterized generator creates brake system instances with $n$ components (total EMF objects $\approx 7n$). Two branches diverge from the base; each contributes $k$ commits modifying disjoint components (no conflicts) to isolate scalability from conflict handling. We report the median over five runs (one warmup discarded) on a 12-core \texttt{aarch64} machine with 8\,GB RAM, OpenJDK~21. All timings reported in this paper, including the Time column of \cref{tab:comparison}, were measured on this machine so that they are directly comparable.

\paragraph{Model size scaling.}
\Cref{tab:perf-model-size} shows merge time as a function of model size with a fixed history of $k=5$ commits per branch.

\begin{table}[t]
\centering
\caption{Median merge time vs.\ model size ($k=5$ commits/branch).}
\label{tab:perf-model-size}
\small
\begin{tabular}{rrrr}
\hline
\textbf{Components} & \textbf{EMF objects} & \textbf{Merge time (ms)} & \textbf{ms/commit} \\
\hline
10   & $\sim$70    & $28.1$  & 5.6 \\
50   & $\sim$350   & $32.2$  & 6.4 \\
100  & $\sim$700   & $44.2$  & 8.8 \\
250  & $\sim$1750  & $88.3$  & 17.7 \\
500  & $\sim$3500  & $196.0$ & 39.2 \\
750  & $\sim$5250  & $277.4$ & 55.5 \\
1000 & $\sim$7000  & $402.5$ & 80.5 \\
\hline
\end{tabular}
\vspace{-0.3cm}
\end{table}

\paragraph{History length scaling.}
\Cref{tab:perf-history} shows merge time as a function of the number of commits replayed, with model size fixed at $n=50$ components.

\begin{table}[t]
\centering
\caption{Median merge time vs.\ history length ($n=50$ comp.).}
\label{tab:perf-history}
\small
\begin{tabular}{rrr}
\hline
\textbf{Commits} & \textbf{Merge time (ms)} & \textbf{ms/commit} \\
\hline
1  & $19.9$  & 19.9 \\
5  & $30.6$  & 6.1 \\
10 & $47.2$  & 4.7 \\
25 & $93.8$  & 3.8 \\
50 & $184.7$ & 3.7 \\
\hline
\end{tabular}
\vspace{-0.5cm}
\end{table}

\paragraph{Interleaving and conflict overhead.}
\Cref{tab:perf-model-size,tab:perf-history} use disjoint branches to isolate scalability from conflict handling. The Time column in \cref{tab:comparison} reports wall-clock merge times for the RQ1 scenarios, which exercise all merge outcomes including dependency graph construction, interleaving, and conflict detection. The pattern is the same in all three case studies: the \textsc{consequential} scenarios are fastest (3.5--4.8\,ms), because a cycle in the dependency graph is detected before any replay occurs, while the scenarios that do replay complete in under 20\,ms. That the three case studies agree to within a few milliseconds per scenario category, despite differing in domain, metamodels, and kind of consistency relation, indicates that merge cost is governed by the size of the change history being replayed rather than by the content of the consistency relations.

\paragraph{Summary (RQ3).}
Merge time scales approximately linearly with model size up to 1000 components, growing monotonically from 28\,ms at $n\!=\!10$ to 403\,ms at $n\!=\!1000$.
With respect to history length, total merge time likewise grows monotonically (20\,ms at $k\!=\!1$ to 185\,ms at $k\!=\!50$) while the marginal per-commit cost falls (from 19.9\,ms/commit at $k\!=\!1$ to 3.7\,ms/commit at $k\!=\!50$), reflecting the amortization of fixed setup costs over longer histories. For the largest configuration tested ($n=1000$ components, $k=5$ commits), the merge completes in under half a second. These results indicate that the replay-based approach is practical for the model sizes and branch histories typical of multi-model development.

\subsection{Threats to Validity}

\paragraph{Internal validity.}
Both approaches operate on identical Git repository states; conflict counts are derived from automated test assertions with per-scenario oracle checks.

\paragraph{External validity.}
The evaluation covers three case studies with three metamodels each, spanning two automotive domains and one software-engineering domain, and two kinds of consistency relation (value propagation in Case~Studies~1 and~2, structural constraints in Case~Study~3). Results may nonetheless not generalize to domains with fewer cross-model dependencies. Two residual threats remain. First, $M_2$ in Case~Study~3 is an abstracted code model rather than a full Java abstract syntax tree; it captures the abstraction level at which the subject project states its own architecture constraints, but a finer-grained code model could exhibit consistency relations we do not exercise. Second, RQ2 and RQ3 are measured on Case~Study~1 only, so the robustness and scalability results are not replicated across domains. No publicly available datasets of real-world multi-model change histories with consistency relations exist; the 130~generated scenarios in RQ2 partially mitigate this limitation.

\paragraph{Construct validity.}
Blocking conflict count is the primary metric: each EMF~Compare conflict corresponds to a model element requiring manual inspection.

%% file: sections/07_relatedwork.tex
\section{Related Work}
\label{sec:relatedwork}

We discuss related work along four dimensions: state-based model merge, operation-based versioning, concurrent synchronization in bidirectional transformations, and multi-model consistency management.

\paragraph{State-Based Model Comparison and Merge.}
State-based tools such as EMF~Compare~\cite{emfcompare,brun2008model} and EMF~Diff/Merge perform three-way merge by comparing model snapshots. Formalizations by Westfechtel~\cite{westfechtel2014merging} and configurable approaches such as E3MP~\cite{sharbaf2020configurable}, EML~\cite{kolovos2006merging}, and DSMCompare~\cite{jafarlou2026domainspecific} support domain-specific conflict detection. However, these approaches operate on model states and treat automatically propagated updates uniformly with user-authored edits, leading to false-positive conflicts when consequential changes arise from consistency rules.

\paragraph{Operation-Based Model Versioning.}
Operation-based approaches usually record edit histories and improve conflict detection~\cite{brosch2012introduction,mens2002survey}. For example, EMF\-Store~\cite{koegel2010emfstore,koegel2010operation} demonstrates reduced conflicts, but operates on single models without cross-model awareness. Formal approaches such as \cite{taentzer2014fundamental,barkowsky2022multi} and hybrid techniques such as AMOR~\cite{brosch2012introduction} express differences between model versions as operations, employ these operations during merging, and address structural conflicts, yet do not consider transformation-triggered consequential changes across multi-models. Sharbaf et al.~\cite{sharbaf2023conflict} confirm that existing merge techniques primarily focus on syntactic conflicts.

A complementary line of work addresses how the conflicts that do reach a developer are resolved: CoMPers~\cite{sharbaf2025compers} provides configurable, personalizable conflict-management strategies for collaborative modeling. This is orthogonal to our contribution rather than an alternative to it. We reduce the number of conflicts a developer is asked to decide, by regenerating consequential changes instead of merging them; strategy-based frameworks determine how the remaining decisions are made. The two compose, and because such strategies do not fully automate resolution, a genuine direct conflict still encodes two incompatible design intents that only a human can reconcile, reducing the number of conflicts at the source retains its value even where they are available.

\paragraph{Concurrent Model Synchronization.}
Concurrent synchronization in bidirectional transformations has been studied for pairs of models~\cite{xiong2013synchronizing,diskin2011state}. TGG-based approaches~\cite{hermann2012concurrent,fritsche2020precedence} detect conflicts via rule dependencies, while multiary consistency restoration across transformation networks is analyzed in~\cite{stevens2020maintaining}. These techniques reconcile overlapping original updates but do not distinguish original from consequential changes propagated through transformation rules.

\paragraph{Multi-Model Consistency Management.}
Transformation networks and their correctness are formalized by Klare~\cite{klare2022building} and realized in \vitruv~\cite{klare2021enabling}. Consistency as a relation over model tuples is studied in~\cite{PascualBeckertUlbrichKirstenPfeifer2024}, while multiary consistency structures are analyzed in~\cite{stunkel2021comprehensive}. Surveys of change metamodels~\cite{weber2025how} indicate that none record provenance distinguishing original from consequential updates, which is required by our merge approach.

\paragraph{Collaborative Multi-Model Infrastructures.}
\vitruv{} is not the only infrastructure that manages several interrelated models for concurrently working engineers, and it is worth being precise about where the difference lies. DesignSpace maintains consistency across multi-domain engineering models by means of reactive links~\cite{ratiu2022reactive}, and its collaboration layer~\cite{herac2023flexible} is, like ours, \emph{operation-based} rather than state-based: model evolution is persisted as change-based persistence over a fixed set of atomic operations (element and property creation, deletion and update), and versions are nodes in a model history tree whose branches are merged by concatenating their operation sequences. It therefore satisfies the first of our two portability requirements, ordered, fine-grained deltas, without modification.

The difference is provenance. Conflict detection there compares the operation sequences of two branches from their common predecessor and reports three cases: the same element deleted on both branches, an element deleted on one branch and a property of it changed on the other, and the same property changed or deleted on both~\cite{herac2023flexible}. These are, notably, the same syntactic situations our direct-conflict detection covers. What the operation model does not carry is a label distinguishing an operation an engineer performed from one a consistency mechanism derived, so every operation in a sequence is weighed equally at merge time; the reported resolution correspondingly selects a winner by a per-user flag rather than by regenerating derived state. Consequential overlaps are therefore not separable from genuine ones, which is the gap our approach addresses rather than a shortcoming of that infrastructure, it is designed for lock-free multi-user collaboration across co-existing metamodels, a different problem.

DesignSpace's consistency mechanism is separately reported: \emph{reactive links}~\cite{ratiu2022reactive} connect individual property values across artifacts from different tools, and carry a propagation operator (assignment, or a comparison that checks a constraint), a direction (bidirectional, unidirectional, or asymmetric), and a trigger (automatic or manual). Automatic triggers support chain propagation, one link evaluation causing another. This is a genuine consistency-preservation mechanism at a granularity comparable to ours, and it satisfies our determinism requirement for the automatic case. The links do react to \emph{where} a change originated, but in the sense of which end of the link was modified, which selects the propagation direction; it is not a record of whether the value was written by an engineer or by a previous link evaluation. Consequently a propagated value is indistinguishable from an authored one in a subsequent merge, which is precisely the distinction our approach consumes.

Openflexo~\cite{hnaini2025openflexo} takes a different route again, \emph{federating} rather than unifying: each model remains in its own technological space and is related to others through a federation model whose concepts carry executable behaviours that check and repair consistency. Consistency preservation is thus expressible, but the prerequisites our merge needs are not met for a different reason: change is detected by comparing duplicated property values rather than recorded as an ordered sequence of deltas, and the authors themselves note that the more direct binding of a property to its source would not support change tracking in their execution model. Without a change history there is nothing to replay, so requirement~(i) fails before provenance becomes relevant.

\paragraph{How common is merge support at all?}

It is worth situating these three tools in the wider landscape, because the premise of this paper is that merge in a consistency-managed multi-model setting is under-served. Choudhury et al.~\cite{choudhury2025landscape} systematically review 32~collaborative MDSE technologies and classify each by capability. Of those 32, only \emph{three} provide a diff/merge engine and five a version-control system; the two most widely used of the three are EMF~Compare and EMF~Diff/Merge, which supports our choice of baseline in \cref{sec:evaluation}. Change provenance does not appear as a dimension of that review's capability model at all, the vocabulary in which collaborative modelling technologies are currently described does not distinguish an authored change from a derived one. We take this as corroboration that the distinction is not merely unimplemented but largely unarticulated, rather than as proof that no tool records it, since a capability survey classifies tools by advertised function rather than by internal representation.

We are therefore careful about the strength of the claim. We do not assert that \vitruv{} is uniquely capable of supporting replay-based merge: reactive links and the DesignSpace operation model both come close, and adding a provenance label to either is a plausible extension rather than a redesign. What we assert is narrower and, we think, more useful: the original-versus-consequential distinction is not recorded by the infrastructures we examined, it is absent from the capability vocabulary of the surveyed landscape, and \cref{sec:formalization:scope} states exactly what a framework must provide to obtain the merge described here.

%% file: sections/08_conclusion.tex
\section{Conclusion and Future Work}
\label{sec:conclusion}

We presented a replay-based three-way merge for multi-model consistency systems that records the original changes of each commit and interleaves them from both branches via a footprint dependency graph, regenerating consequential changes through the consistency engine rather than merging them. A provenance-aware conflict classification distinguishes direct conflicts (original vs.\ original) from consequential conflicts (cycles in the dependency graph).

We formalized the merge operator and its footprint dependency graph (\cref{sec:formalization:interleaving}), showed that successful merges preserve
both branches' original intent, and implemented the approach as an extension to \vitruv. Evaluation on three case studies (brake systems, automated driving, and code--architecture consistency in the TEAMMATES open-source project) across 14 scenarios shows that our approach reports 10 blocking conflicts where EMF~Compare reports 37, as consequential overlaps are resolved automatically by the ordering algorithm. The third case study maintains structural consistency relations rather than value propagation, indicating that the reduction does not depend on the consistency relation being arithmetic.

\paragraph{Limitations and future work.}
The formalization assumes fixed intra-branch commit order; relaxing this by modeling intra-branch dependencies explicitly could resolve additional consequential overlaps. Fixed consistency rules are assumed; evolving rules would require merging rule definitions. While framework-independent in principle, the prototype is coupled to \vitruv's provenance-aware change metamodel; extending provenance support to other delta-based systems~\cite{weber2025how} would enable broader applicability. Future work includes validation on larger multimodels, post-merge metamodel validation, and interactive conflict resolution~\cite{jafarlou2026domainspecific}.